\documentclass{aa}  

\usepackage{graphicx}
\usepackage{txfonts}
\usepackage{todonotes}

\begin{document}

   \title{Observed binary populations reflect the Galactic history}
    
   \subtitle{Explaining the orbital period-mass ratio relation in wide hot subdwarf binaries.}
    

   \author{J. Vos\inst{1}\thanks{E-mail: joris.vos@uv.cl}
          \and
          A. Bobrick\inst{2}
          \and
          M. Vu\v{c}kovi\'{c}\inst{3}
          }

   \institute{Institut f\"{u}r Physik und Astronomie, Universit\"{a}t Potsdam, Karl-Liebknecht-Str. 24/25, 14476, Golm, Germany
         \and
        Lund University, Department of Astronomy and Theoretical physics, Box 43, SE 221-00 Lund, Sweden
        \and 
        Instituto de F\'{\i}sica y Astronom\'{\i}a, Universidad de Valpara\'{\i}so, Gran Breta\~{n}a 1111, Playa Ancha, Valpara\'{\i}so 2360102, Chile
              }

   \date{}

  \abstract
   {
   Wide hot subdwarf B (sdB) binaries with main-sequence companions are outcomes of stable mass transfer from evolved red giants. The orbits of these binaries show a strong correlation between their orbital periods and mass ratios. The origins of this correlation have, so far, been lacking a conclusive explanation.
   }
   {
   We aim to find a binary evolution model which can explain the observed correlation.
   }
   {
   Radii of evolved red giants, and hence the resulting orbital periods, strongly depend on their metallicity. We performed a small but statistically significant binary population synthesis study with the binary stellar evolution code MESA. We used a standard model for binary mass loss and a standard metallicity history of the Galaxy. The resulting sdB systems were selected based on the same criteria as was used in observations and then compared with the observed population.
   }
   {
   We have achieved an excellent match to the observed period-mass ratio correlation without explicitly fine-tuning any parameters. Furthermore, our models produce a very good match to the observed period-metallicity correlation. We predict several new correlations, which link the observed sdB binaries to their progenitors, and a correlation between the orbital period, metallicity, and core mass for subdwarfs and young low-mass helium white dwarfs. We also predict that sdB binaries have distinct orbital properties depending on whether they formed in the Galactic bulge, thin or thick disc, or the halo.
   }
   {
   We demonstrate, for the first time, how the metallicity history of the Milky Way is imprinted in the properties of the observed post-mass transfer binaries. We show that Galactic chemical evolution is an important factor in binary population studies of interacting systems containing at least one evolved low-mass ($M_{\rm init} < 1.6\,M_{\odot}$) component. Finally, we provide an observationally supported model of mass transfer from low-mass red giants onto main-sequence stars.
   }
  %

   \keywords{binaries: spectroscopic, Stars: evolution, Stars: mass-loss, subdwarfs, Galaxy: evolution}

   \maketitle
%

\section{Introduction}

Hot subdwarf-B (sdB) type stars are core-helium-burning stars that have lost the majority of their hydrogen-rich envelope ($M_{\rm env} \approx 0.01\,M_{\odot}$).  Most of these systems have masses very close to the core-helium-flash mass of $0.47\,M_{\odot}$ \citep{Heber2009, Heber2016}. Although many different formation channels have been proposed, it is currently accepted that all sdB stars form solely due to binary interactions \citep{Heber2016}.

Hot subdwarf binaries are excellent probes of binary evolution. Since the masses of the sdB stars in these binaries are typically close to the core-helium-flash mass,  many of them must have formed near the end of the red giant phase, which in turn constrains their progenitor properties. Furthermore, the long-period sdBs with main-sequence (MS) companions, which are the subject of this study, are believed to have undergone only one stable mass transfer (MT) phase, and thus may be used to constrain the properties of mass transfer. Finally, these binaries are typically double-lined spectroscopic sources, which allows one to study both the sdBs and their cool companions.

There are three main formation channels for sdB stars. In the common envelope (CE) ejection channel, the sdB star forms as the core of an evolved red giant which has lost most of its hydrogen-rich envelope in a common envelope episode \citep{Paczynski1976, Han2002}.  The CE ejection channel produces sdB binaries with short orbital periods (i.e. hours to a few tens of days) and white dwarf (WD) or MS companions. Conversely, in the stable Roche-lobe overflow (RLOF) channel, a red giant loses its envelope through stable mass transfer \citep{Han2000, Han2002}.  The stable RLOF channel produces wide sdB binaries with orbital periods of the order of years, up to about $1600$ days. Finally, in the WD merger channel, the sdB star forms as a result of a merger of two white dwarfs \citep{Webbink1984}. The WD merger channel creates single sdB stars with masses that can potentially be higher than the canonical value. The outcomes of the different sdB formation channels were studied through binary population synthesis by \citet{Han2003} and, more recently, by \citet{Chen2013}. In this study, we focus on long-period sdB binaries with MS companions which formed through the stable RLOF channel and ignite degenerately through a helium flash.

Composite long-period sdB binaries have been predicted by \citet{Lamontagne2000}, \citet{Green2001}, \citet{Aznar2001} and \citet{Reed2004}, but the first solved long-period systems based on spectroscopic monitoring programmes were published only a few years ago \citep{Oestensen2011, Deca2012}. Recently, the orbital properties of 23 wide sdB binaries have been published \citep[see, for example,][for an overview]{Vos2019}. 

These observations have already led to improvements in the binary interactions models for stable RLOF from red giants. First binary population studies of long-period sdB binaries predicted orbital periods of the order of a few hundred days \citep{Han2002}. The periods of the first observed systems with solved orbits, however, were significantly longer. These longer periods were explained with a more detailed treatment of the angular momentum loss and the inclusion of atmospheric RLOF \citep{Chen2013}. A second discovery was that almost all long-period sdBs had eccentric orbits, which contradicted the tidal circularisation models that predict that all sdB progenitor binaries should circularise long before they start transferring mass. Using the stellar evolution code Modules for Experiments in Stellar Astrophysics \citep[MESA,][]{Paxton2011, Paxton2013, Paxton2015, Paxton2018, Paxton2019}, \citet{Vos2015} proposed two possible mechanisms that could explain the observed eccentricities: phase-dependent mass loss and the interaction of a circumbinary disc with the binary. 

The most recent discovery was the strong relation between the mass ratio and the orbital period in these wide sdB binaries. This relation was attributed to the stability of RLOF \citep{Vos2019}, but up to now, this has not been explained by binary population synthesis models. In this study, we show that the observed relation between the mass ratio and the orbital period in long-period sdB binaries can be fully explained by the chemical evolution of our Galaxy.

Galactic evolution has a strong effect on the period distribution of long-period sdBs. The Galactic metallicity, [Fe/H],  evolves with time, and at the same time metallicity has a strong effect on the sizes of the red giant progenitors of sdB stars. Stars more massive than $1.5\,M_\odot$ formed relatively recently and, therefore, produce sdB binaries at solar metallicity. Stars of about $0.9\,M_\odot$ formed early in the Galactic history and, therefore, produce sdB binaries at sub-solar metallicities of about $-0.4$ \citep{Robin2003}. Compared to solar metallicities, the red giant radii of such stars are about $30\,\%$ smaller \citep{MIST}, and the resulting sdB periods are, correspondingly, shorter. Therefore, to model the $P\,-\,q$ relation of the whole population of sdB binaries, one needs to model how the Galactic metallicity and star formation rate evolved with time.

In this paper, we perform a population synthesis of composite sdB binaries by using the binary stellar evolution code MESA, by adopting a conventional model of red-giant mass transfer and by accounting for the Galactic chemical evolution by using the Besançon model \citep{Robin2003,Robin2014,Czekaj2014}. Compared to typical population studies which ignore the Galactic evolution, we have been able to reproduce the observed $P\,-\,q$ distribution with minimal assumptions and without explicitly fine-tuning any parameters.


\section{Observations}
\label{sec:Obs}
The current sample of long-period hot subdwarf binaries contains 25 systems with solved orbital periods, of which 23 systems have a mass ratio. This sample contains six systems that were part of the long-term monitoring programme with the Mercator telescope \citep{Vos2012, Vos2013, Vos2017}. Two systems were solved by \citet{Barlow2013} based on observations with the Hobby-Eberly telescope. For the original long-period sdB binary solved by \citet{Deca2012}, we used updated orbital parameters from \citet{Deca2018}, and 11 binaries are part of the UVES survey \citep{Vos2019}. Furthermore, the sample contains one system that has an orbital period determined from lightcurves \citep{Otani2018}, but no mass ratio can be determined this way. Recently, also two long-period binaries with sdO components were analysed \citep{Molina2020}. The two sdO stars have masses consistent with the canonical sdB mass, and are likely in a later evolutionary stage: He-shell burning. As the orbital parameters are not affected by the evolution from He-core burning to He-shell burning, these systems are also included in this sample.
Most of these systems were selected based on GALEX colours, and the selection criteria of the UVES and Mercator sample are discussed in \citet{Vos2018}.  

Before comparing the observations with models, it is important to check if the selection criteria of the sample and detection limits of the telescopes and instruments have any impact on the observed correlations. These effects have been studied in detail by \citet{Vos2019}, where they found that neither the target selection nor the sensitivity of the telescope could be responsible for the observed correlations in orbital period with mass ratio. The sensitivities of the instruments, UVES and HERMES, are also sufficiently high to detect systems with orbital period significantly longer than those observed in this sample. The sample that we analyse here can, therefore, be considered to have no bias for the orbital features that we aim to explain. At the same time, as the observational sample is magnitude-limited to the bright systems in the solar neighbourhood, we expect it to contain significantly more systems from the Galactic thin disc, rather than the sparse and less massive thick disc or the halo. The sample contains no systems from the distantly-located bulge as that region was explicitly avoided.

The wide sdB binary sample shows two main correlations. The first one is the correlation between the orbital period and eccentricity, and the second one is the correlation between the orbital period and the mass ratio. The observed eccentricity of these systems is unexpected as they should have tidally circularised before the onset of mass transfer. This unexplained eccentricity is also observed in other evolved binaries, such as, for example, barium stars, post-AGB stars and He-WD binaries \citep[see][for an overview]{Vos2017}. \citet{Vos2015} found that phase-dependent mass-loss and circumbinary discs can explain the observed eccentricity in wide sdB binaries. However, while their models allowed for the observed parameters, they could not predict the observed correlation with the orbital period.

In this work, we focus on the relation between the orbital period and the mass ratio. In the correlation, longer orbital periods ($P$) correspond to smaller mass ratios $q=M_{\rm sdB}/M_{\rm comp}$. Assuming that the sdB binaries formed through degenerate ignition from low-mass progenitors (M $\lesssim 2.0 M_{\odot}$) and that the sdB mass is close to canonical ($0.47\,M_{\odot}$), this $P\,-\,q$ relation is equivalent to saying that longer orbital periods correspond to larger companion masses. The plots showing the observed orbital periods and mass ratios are presented in Section \ref{sec:ResMain}.

An interesting notion in the observations is that there is a secondary group of systems that follow the same $P\,-\,q$ and $P\,-\,e$ correlations as the main group but at shorter orbital periods. In observable features, there is no difference known between hot subdwarfs in the main group and the secondary group. There are only a few systems found in the secondary group, which makes a statistical comparison difficult. Up to now, the reason for the difference in both groups is unknown, although we discuss the possible reasons in Section~\ref{sec:DiscStellar}.


\section{Binary evolution model}
\label{sec:BinEvol}
\begin{figure}
    \includegraphics[width=\linewidth]{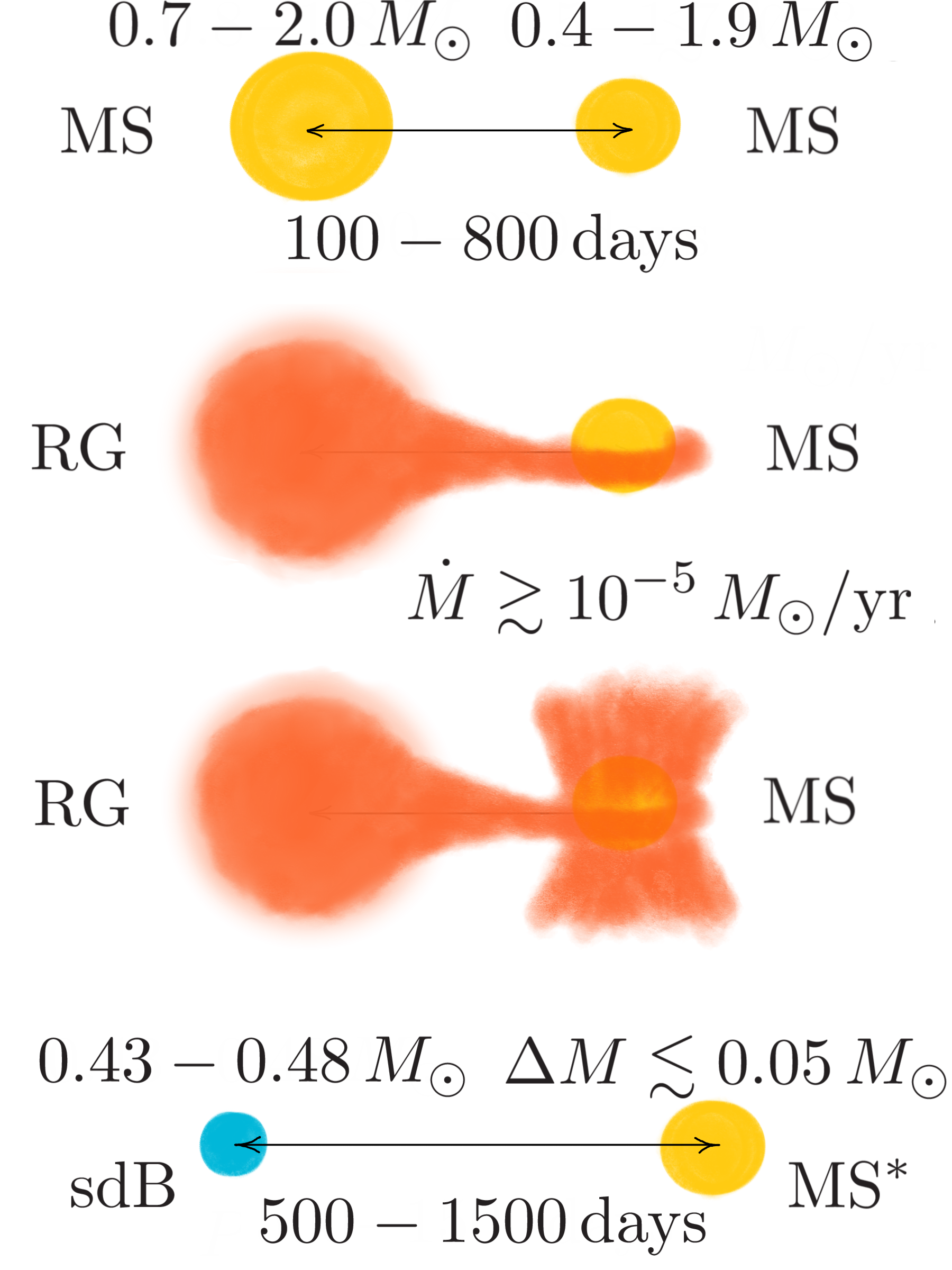}
    \caption{Formation channel for the main $P\,-\,q$ branch of wide-period composite sdB binaries. Initially, a $0.7\,M_\odot$--$2.0\,M_\odot$ main-sequence primary (MS) evolves to become a developed red giant (RG), which initiates mass transfer through Roche lobe overflow. The mass transfer rate gradually grows and exceeds $10^{-5}\,M_\odot/{\rm yr}$, at which point the accretor cannot accept the incoming mass and ejects it in an outflow (shown schematically). Even though the red giant loses all its envelope, its degenerate core ignites helium and turns into a luminous sdB star. The companion main-sequence star accretes only a small amount of mass of $\lesssim 0.05\,M_\odot$, but, nevertheless, becomes polluted (MS$^*$). The radius of the RG is strongly sensitive to the metallicity of the primary, which in turn correlates with its age and mass due to Galactic evolution. This way, the Galactic age-metallicity correlation is an important factor determining the final periods of long-period sdB binaries.}
    \label{fig:ChannelDiagram}
\end{figure}

\begin{figure}
    \includegraphics[width=\linewidth]{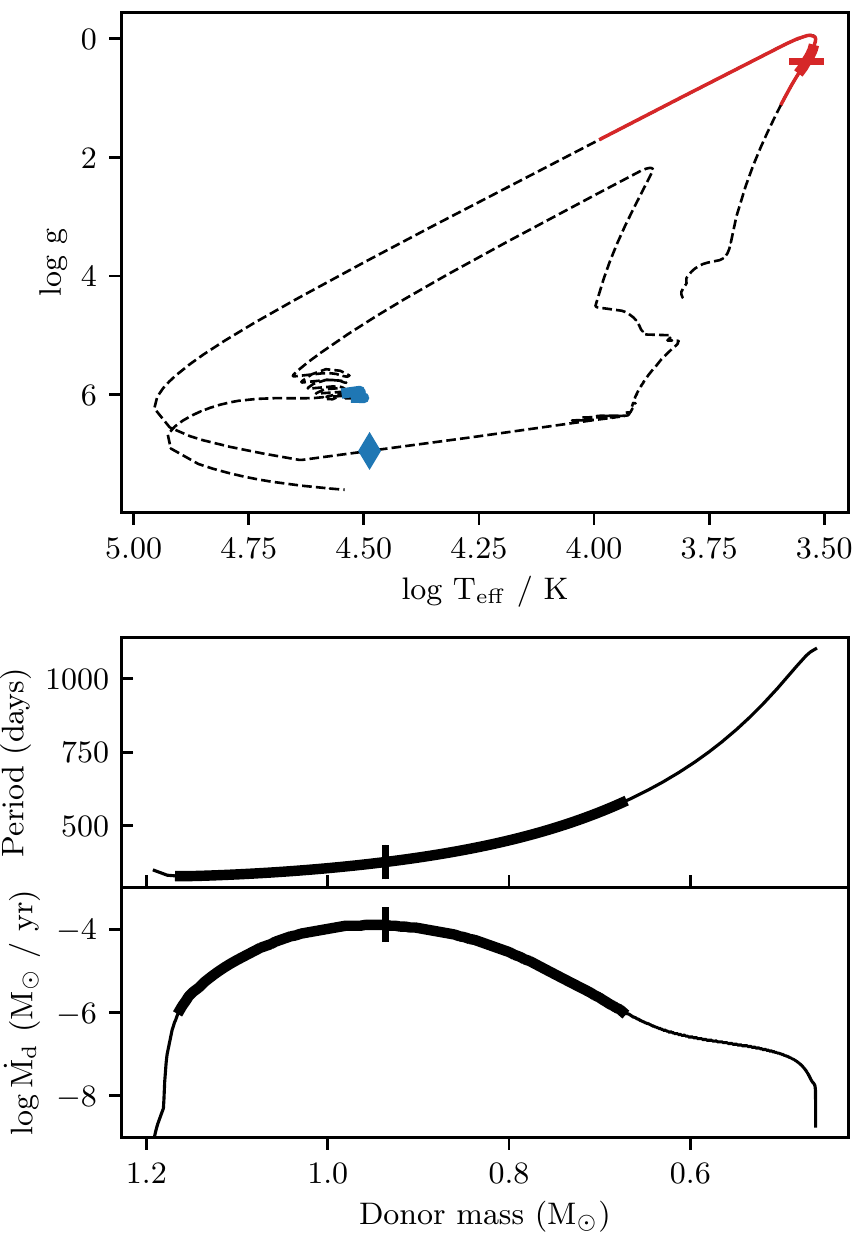}
    \caption{An example MESA track for a binary with a 1.2\,$M_{\odot}$ donor star, a 0.85\,$M_{\odot}$ companion, an initial period $P_{\rm i}$ = 350 days and a metallicity ${\rm [Fe/H]} = -0.15$, which produces a hot composite sdB binary. 
    The top panel shows the track for the donor star in the HR diagram. The mass loss phase is marked in red; the thin red line shows the phase when $\dot{M} > 10^{-9}\,M_\odot/{\rm yr}$, thick red line corresponds to $\dot{M} > 10^{-6}\,M_\odot/{\rm yr}$ and the moment of maximum mass loss is marked with a horizontal red marker. The first He flash is marked with a blue diamond and the core He burning phase is shown in a thick blue line. The phase between the He flash and the core He burning is very short compared to the core He burning phase. The bottom panel shows the orbital period and the mass loss rate during the mass loss phase. The same markings are used as in the HR diagram: the thin black line shows the phase with $\dot{M} > 10^{-9}\,M_\odot/{\rm yr}$, thick black line corresponds to $\dot{M} > 10^{-6}\,M_\odot/{\rm yr}$ and maximum mass loss is marked with a vertical black marker.}\label{fig:example_mesa_model}
\end{figure}

Long-period sdB binaries result from stable mass transfer from evolved red giants onto main-sequence stars. To result in an sdB object, the naked helium core of the red giant has to ignite helium and have less than $0.01\,M_\odot$ of hydrogen remaining in its atmosphere \citep{Heber2016}.

We adopt a standard observationally-motivated formation scenario for long-period sdB binaries and schematically show it in Fig.\,\ref{fig:ChannelDiagram}. Additionally, in Fig.\,\ref{fig:example_mesa_model}, we show a typical evolutionary MESA track leading to a long-period sdB binary, starting from a 1.2 $M_{\odot}$ donor star. We consider binaries with primary masses between $0.7\,M_\odot$ and $2.0\,M_\odot$ and initial periods between $100$ and $800$ days. These are primaries which may have left the main sequence by the present time, ignited their helium core degenerately \citep{Hurley2000} and initiated mass transfer near the end of the red-giant branch \citep{Han2002,Chen2013}. Initially eccentric, these binaries circularise before the mass transfer starts \citep{Vos2015, Beck2020}. As the red giant expands, it fills its Roche lobe and initiates the mass transfer. For a broad range of initial binary mass ratios, $q_i = M_{\rm primary}/M_{\rm comp}\lesssim 2.0$, the mass transfer proceeds stably at mass transfer rates below $10^{-3}\,M_\odot/\textrm{yr}$, resulting in long-period binaries \citep{Tauris2000, Pavlovskii2015}. The exact stability criterion for mass transfer also depends on the evolutionary stage of the red giant and the mechanism of mass-loss in the binary.

Observations of companions of long-period sdB binaries indicate that they accrete only a small, up to a few times $0.01\,M_\odot$, amount of material, that is to say they form through non-conservative mass transfer \citep{Vos2018}. Therefore, we adopt the simplest non-conservative mass transfer scenario consistent with these observations. As the red giant donor expands, progressively larger parts of its envelope end up outside the Roche lobe. Gravitationally unbound from the donor, these layers flow through the $L_1$ point and form an accretion disc. Initially, mass transfer rates are low, and we assume that most of the transferred mass is conservatively accreted on the companion. As the mass transfer rates reach $\dot{M}\gtrsim 10^{-5}-10^{-6}\,M_\odot/\textrm{yr}$, the companion accretes a few $0.01\,M_\odot$ and gets critically spun up by accretion \citep{Popham1991}. Subsequently, the companion cannot accrete any further material as the material cannot efficiently remove the excess angular momentum in order to land on the companion \citep{Popham1991, Paczynski1991,Deschamps2013}.  Additionally, at similar mass transfer rates, the gravitational energy released due to accretion cannot be efficiently radiated by the accretor, making it swell, which is also expected to lead to mass loss \citep{Kippenhahn1977,Pols1994,Toonen2012}. As we discuss in Section~\ref{sec:SimMain}, both effects occur at similar mass transfer rates, but our MESA simulations suggest that typically over-spinning takes place first. 

We assume that mass gets lost through a non-collimated jet-like outflow with the angular momentum of the accretor, see for example \citet{Tauris2006} and \citet{Shiber2018}: 
\begin{equation}
\dfrac{\dot{J}_z}{J_z}=\dfrac{(M_{\rm RG}/M_{\rm comp})^2}{1+(M_{\rm RG}/M_{\rm comp})}\dfrac{\dot{M}_{\rm RG}}{M_{\rm RG}}    .
\end{equation}
As follows from \citet{Chen2013} and as we discuss in detail in Section~\ref{sec:DiscStellar}, the exact choice of the angular momentum loss prescription should only have a small effect on the final period-mass ratio correlation of sdB binaries. Depending on the binary parameters, mass transfer rates may reach values of about $10^{-2}\,M_\odot/\textrm{yr}$. In this case, we assume that binaries enter a common envelope stage, for example \cite{Pavlovskii2015}, and produce short-period systems.

The final periods of wide sdB binaries are sensitive to their metallicity, for example \cite{Chen2013}. At the same time, Galactic metallicity has evolved with time. The oldest stars generally have metallicity of about $0.4\,\textrm{dex}$ lower than the presently-formed stars \citep{Edvardsson1993}. We illustrate the importance of metallicity through a simple analytic argument here, and present a detailed demonstration based on MESA runs further in Section~\ref{sec:effect_initial_parameters}. Let us consider two binaries: Binary 1 has primary mass of $1.0\,M_\odot$ and metallicity ${\rm [Fe/H]} = -0.4$, Binary 2 has the same primary mass of $1.0\,M_\odot$, but its metallicity is solar (${\rm [Fe/H]} = 0$). Binary 1 represents an average system (for its primary mass) which formed about $8.4\,\textrm{Gyr}$ ago in the Galaxy and reached the tip of the red-giant branch (RGB) today. The primary then has a radius of about $135\,R_\odot$ at the tip of the RGB \citep{MIST}. Binary 2 represents the same system as Binary 1, assuming it was born at solar metallicity, a much less typical metallicity for the Galaxy at the time the binary was born. Due to its overestimated metallicity, Binary 2 would have a radius of $165\,R_\odot$ at the tip of the RGB.

We can now estimate the effect of metallicity on the final periods of these binaries. For both Binary 1 and Binary 2 we assume a $0.7\,M_\odot$ companion which, according to our MESA simulations described later, could lead to the production of an sdB from the primaries in both cases. Due to the difference in red giant radii, the initial periods of these binaries differ by 35 per cent. We can assume that the ensuing mass transfer is fully non-conservative \citep{Vos2018} and that the system loses mass with the angular momentum of the accretor. The final periods after the mass transfer phase are then given by \cite{Soberman1997}:
\begin{equation}
\label{eq:PFinNonCons}
    \dfrac{P_{\rm sdB}}{P_{\rm init}}=\left(\dfrac{M_{\rm RG}}{M_{\rm sdB}}\right)^3\cdot\left(\dfrac{M_{\rm RG}+M_{\rm comp}}{M_{\rm sdB}+M_{\rm comp}}\right)^2\cdot e^{3\left(\dfrac{M_{\rm sdB}}{M_{\rm comp}}-\dfrac{M_{\rm RG}}{M_{\rm comp}}\right)}
\end{equation}
Assuming a canonical sdB mass of $0.47\,M_\odot$ \citep{Heber2009, Heber2016}, both binaries reach similar final mass ratios of $q$ of about $0.65$. Their periods, however, are $1100\,\textrm{d}$ and $1500\,\textrm{d}$, correspondingly. While the binary with the correctly initialised metallicity ends up on the main branch of the $P\,-\,q$ relation, the period of the binary with solar metallicity is inconsistent with the main branch of the $P\,-\,q$ relation. 

The argument above, consistent with the detailed MESA analysis in Section~\ref{sec:ResMain}, shows that the choice of initial metallicities can affect the final periods by several hundred days. In contrast, since mass transfer is non-conservative in our model, the choice of metallicity should not have any significant direct effect on the final mass ratios $q$. It should also be noted that equation \ref{eq:PFinNonCons} applies to real systems only if we know whether the initial binary does indeed produce an sdB-MS binary. In particular, as shown by \citet{Chen2013}, the final periods of sdBs depend on core mass and the radius of the red giant donor at the end, rather than at the beginning of mass transfer. However, since the radius of red giants even after mass transfer is sensitive to metallicity, the same argument still holds. Generally, in order to produce an sdB, the primary must ignite the core and have lost most of its hydrogen envelope. These conditions may be fully captured only through detailed binary simulations we describe further.

\section{Galaxy evolution model}
\label{sec:GalModel}
An accurate evolutionary model for the Galactic metallicity is needed to generate the present-day population of long-period sdB binaries. The lowest-mass stars reaching the red giant branch at present formed $10\,\textrm{Gyr}$ ago at low metallicities of about $-0.4$ \citep{Edvardsson1993}. The more massive stars of $2.0\,M_\odot$ formed in the last few $\textrm{Gyr}$ at solar-like metallicities. Since the metallicity has a strong effect on the radius of red giants, as we show in Section~\ref{sec:BinEvol}, the history of metallicity also has a strong effect on the shape of the $P\,-\,q$ relation for long-period sdB binaries. In contrast, typical binary population synthesis studies, of sdB binaries and other types of binaries in the Galaxy, have generated populations either by assuming a constant or a uniformly distributed metallicity \citep[see e.g.][]{Izzard2009, Hamers2013, Wijnen2015, Stanway2018}. Although see also for example \citet{Lamberts2019}, \citet{Boco2019} and \citet{Olejak2020}, where the metallicity evolution is accounted for.


\begin{table}
\begin{center}
\caption{The model of the Galaxy used in this study. The columns describe the name, age intervals and the metallicities for the stars in each Galactic population bin. The total stellar mass of the Galaxy was set to $6.43\cdot 10^{10}\,M_\odot$ \citep{McMillan2011}.}
\label{Tab:GPop}
\begin{tabular}{ |c|c|c|c|} 
 \hline
 Population bin & Age, Gyr & Mass fraction & $[{\rm Fe}/{\rm H}]$ \\ 
 \hline
 Thin Disc 1 &$0-0.15$ & $0.030$ & $0.01\pm 0.12$\\
 Thin Disc 2 &$0.15-1$ & $0.069$ & $0.03\pm 0.12$\\ 
 Thin Disc 3 &$1-2$ & $0.076$ & $0.03\pm 0.10$\\ 
 Thin Disc 4 &$2-3$ & $0.072$ & $0.01\pm 0.11$\\ 
 Thin Disc 5 &$3-5$ & $0.132$ & $-0.07\pm 0.18$\\ 
 Thin Disc 6 &$5-7$ & $0.126$ & $-0.14\pm 0.17$\\ 
 Thin Disc 7 &$7-10$ & $0.171$ & $-0.37\pm 0.20$\\ 
 Bulge & $8-10$ & $0.192$ & $0.00\pm 0.40$\\
 Thick Disc & $10$ & $0.123$ & $-0.78\pm 0.30$\\ 
 Halo & $14$ & $0.008$ & $-1.78\pm 0.50$\\ 
 \hline
\end{tabular}
\end{center}
\end{table}

We model the evolution of the Galactic metallicity by using the Besanćon model of the Galaxy \citep{Robin2003}. The Besanćon model describes the evolution of star formation and metallicity in the Galaxy and is calibrated by large photometric surveys of the field. 

The model represents the Galaxy through $10$ population bins: $7$ bins for the thin disc and one bin each for the bulge, thick disc and the halo, all of which had a relatively short star-formation history. Each population bin is associated with its stellar mass fraction and mean and $1$-$\sigma$ range of metallicities. 

We summarise the parameters of the Galactic bins in our study in Table~\ref{Tab:GPop}. In the table, the mass fractions are expressed in terms of the Galactic stellar mass, for which we use the value of $M_{\rm MW} = 6.43\cdot 10^{10}\,M_\odot$ from \citet{McMillan2011}, consistent with the recent study by \citet{Licquia2015}. Thin disc parameters, and the stellar mass fraction of the thick disc are based on the original study by \citet{Robin2003}. We obtain mass fractions for the thin disc bins by integrating the Galactic profiles from \citet{Robin2003}. For the mass fraction of the Galactic bulge and the age of the thick disc, we use more recent data from \citet{Robin2012, Robin2014}. Stars are assumed to form at a constant rate within each stellar bin, and the initial mass function (IMF) is based on Kroupa \& Haywood v6 IMF from \citet{Czekaj2014}.

\section{Simulations} \label{sec:SimMain}

\subsection{Modifications to the MESA code}


In this work, we have used the stellar and binary evolution code MESA \citep{Paxton2011, Paxton2013, Paxton2015, Paxton2018, Paxton2019}, version r10390. In order to follow through the evolution described in Section~\ref{sec:BinEvol}, we have placed a limit on the mass accretion of the secondary based on its rotational velocity. Our MESA models have shown that the companion reaches critical rotational velocity long before it starts swelling due to the thermal energy released from accretion. By this point, the donor loses at most $\lesssim 0.03\,M_\odot$, which is close to the estimates of the mass accreted by the companions of the observed sdBs, for example \citet{Vos2018}. Therefore, the companion is allowed to accrete all mass lost from the donor star until it reaches its critical rotational velocity. At that point, the remainder of the material is considered to be lost from the system with the angular momentum of the accretor. In terms of stability, a system is considered stable as long as the donor mass-loss rate does not exceed $10^{-2}\,M_{\odot}/{\rm yr}$, which is a common criterion in binary evolution studies, for example \citet{Pavlovskii2015}. This is the case of all the long-period hot subdwarf systems produced in our model sets, and we additionally kept track of the peak mass transfer rates reached in our runs. To be clear, we have not used the fixed mass-loss fractions provided in MESA, but dynamically calculate the amount of mass accreted by the companion and lost from the system at each timestep. To extract parameters of interest from the MESA runs we have used the Neural Network assisted Population Synthesis code\footnote{NNaPS is available on GitHub: https://github.com/vosjo/nnaps} \citep[NNaPS,][]{Vos2020nnaps}.

\subsection{Extracting observational properties from the MESA runs}\label{sec:ObsCond}


The outcomes of MESA simulations must be identified with sdB binaries. The term hot subdwarf star was originally a purely spectroscopic classification. More recently, it is mostly used to describe only the core He-burning stars that match the spectroscopic parameters. Therefore, in this study, we identify the MESA outcomes as sdB binaries based on their position on the HR diagram. Specifically, we require the star to be core He-burning, and base the selection criterion on the average effective temperature during the core He-burning phase. If the star reaches an effective temperature between 15\,000 and 40\,000 K during its core He-burning phase, it is considered a hot subdwarf. We also only considered hot subdwarfs that ignite He under degenerate conditions. 

\begin{figure}
    \includegraphics{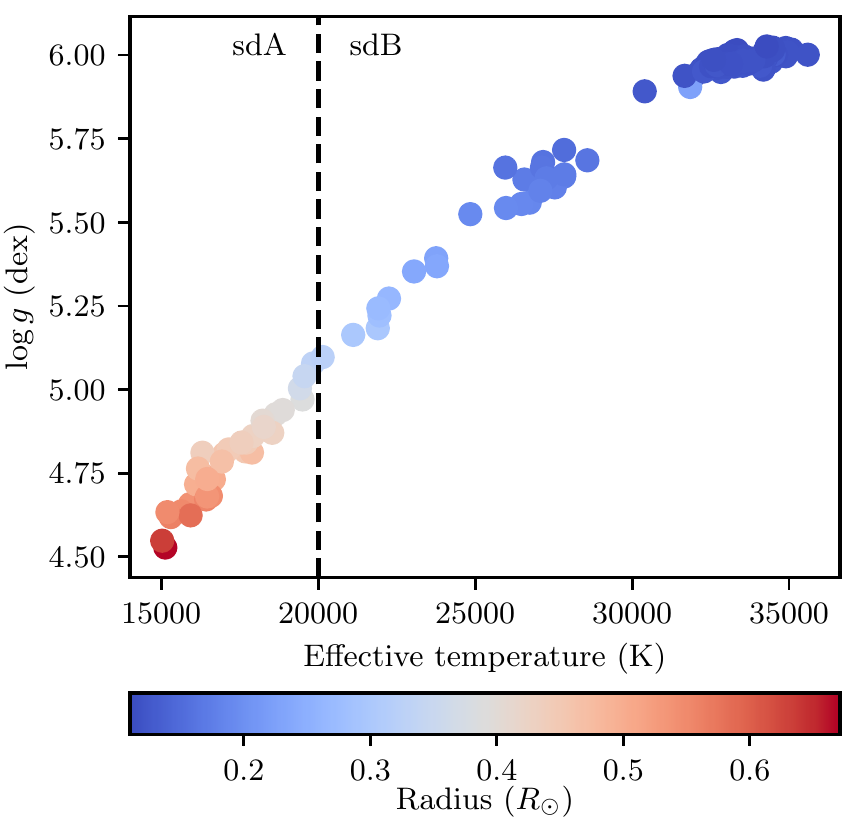}
    \caption{The average effective temperature and surface gravity during the core He burning phase for the hot subdwarfs in the MESA sample. There is a strong relation between $\log{g}$ and $T_{\rm eff}$ for hot subdwarfs formed from progenitors with degenerate He cores. Hot subdwarfs with $T_{\rm eff} <  20\,000\,{\rm K}$ are classified as sdA stars while the hotter subdwarfs are classified as sdB stars.}
    \label{fig:teff_logg_Hecoreburning}
\end{figure}

Furthermore, we make a division between sdA-type stars with $T_{\rm eff} < 20\,000\,K$ and sdB-type stars with higher effective temperatures. For the hot subdwarfs that result from degenerate He-core ignition, the average effective temperature and surface gravity during the He-core burning phase are strongly related. This relation may be seen in Fig.\,\ref{fig:teff_logg_Hecoreburning}, which shows the He-core burning systems resulting from our runs. This figure also shows that the sdBs seem to group together at roughly $28\,000\,{\rm K}$ and at $34\,000\,{\rm K}$. This is a known effect that can be attributed to the treatment of convection during the He-flashes in MESA. The group with the hottest sdBs with temperatures above $34\,000\,{\rm K}$ ignite late on the WD cooling track and convectively mix H and He in the outer layers, igniting most of its hydrogen in a H-flash following shortly after the He-flash, and this way leaving a hot core with almost no H remaining. The systems with lower temperatures experience He-flash early on without igniting hydrogen, this way leaving cooler H atmospheres. See for example \citet{Xiong2017} and \citet{Wu2018} for a discussion. The remaining envelope mass during the core He burning phase for sdB stars ranges from $0.0005$ to $0.004\,M_{\odot}$ while sdA stars have a slightly larger envelope masses ranging from $0.003$ to $0.018\,M_{\odot}$.

Having identified hot subdwarf stars in the MESA models, we select those binaries which would have been identified spectroscopically as hot composite sdB binaries, thus modelling the observed sample. In other words, we apply the same selection criteria to the model sample as the ones used to select the observational composite hot subdwarf sample. Firstly, the observed sample consists only of systems from the thin and thick disc, and potentially from the halo. The systems in the bulge would be too faint for the current observing programmes. The second criterion is that the system should be recognizable as a composite binary. Namely, it should be possible to identify both the sdB and the cool companion in the spectrum. In the observational sample, systems are selected based on visual and IR photometry. However, if it is not possible to recognise both components in the optical spectrum, the systems are rejected from the observational sample. 

\begin{figure}
    \includegraphics{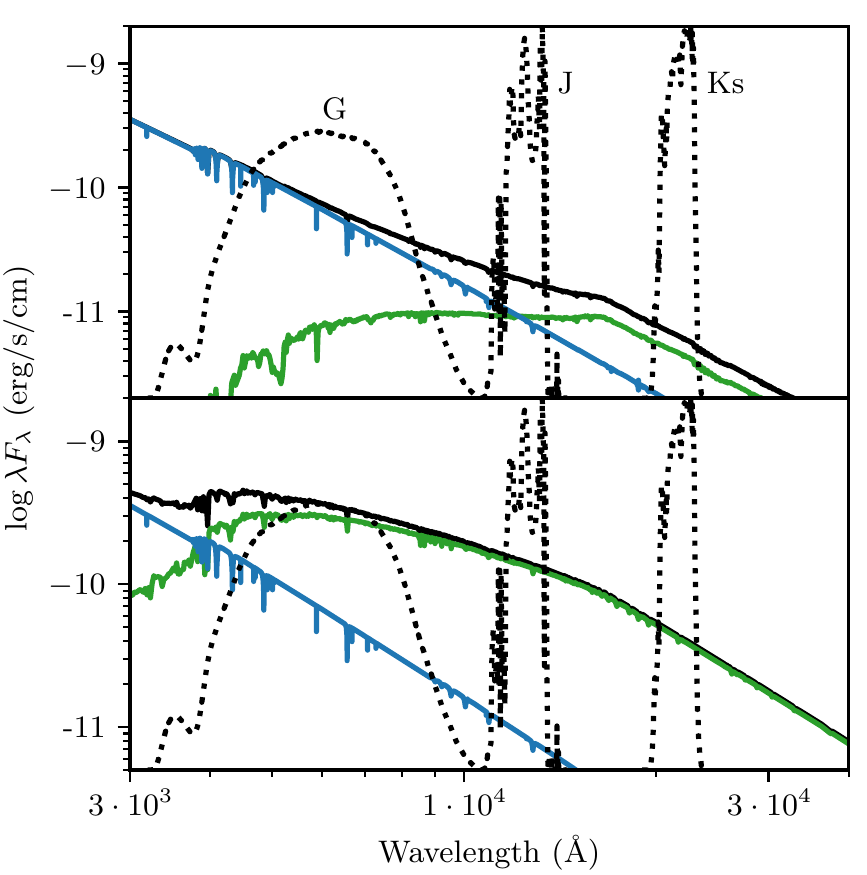}
    \caption{The spectral energy distribution for an sdB+MS binary with 90\% light contribution of the sdB in the Gaia-G band in the top panel, and with 20\% contribution in the bottom panel. The binary SED is plotted in black while the contributions of the sdB and MS component are shown respectively in blue and green. In black-dotted line, the transmission curves for the Gaia-G, 2MASS-J and 2MASS-Ks filter are shown from left to right.}
    \label{fig:visibility_criterion_sed}
\end{figure}

The Gaia catalogue \citep{Gaia2016, Gaia2017} provides the most comprehensive database or stellar positions, magnitudes and proper motions potentially suitable for selecting large samples of candidate long-period composite sdB binaries. It has proven very useful to build catalogues \citep[e.g.][]{Geier2019, Pelisoli2019} and is therefore ideal as a base for an observability criterion. 
We have calculated synthetic colours and spectra for different sdB+MS combinations and found that a criterion based on the sdB flux contribution in the Gaia-G band, as described below, serves as a good indicator for composite binaries. For systems where the sdB flux contribution is between $20$ and $90\,\%$, both components will most likely be recognizable in the optical spectrum. These tests were done with solar metallicity spectra for the companion, and low rotational velocities ($v \sin{i} = 10\,{\rm km}\cdot{\rm s}^{-1}$). Changes in metallicity and rotational velocity can change these results, but we are confident that for this purpose the Gaia-G band selection criterion is sufficient. An example of two sdB+MS systems at both ends of the visibility selection criterion is given in Fig.\,\ref{fig:visibility_criterion_sed}. The systems that fulfil this flux-based selection criterion will be referred to as composite binaries, while the systems that do not fulfil this criterion will be called single-lined binaries (regardless of which component is visible).

\subsection{Main set of runs}
\label{sec:MainSetInit}


We summarise the main and the additional sets of runs performed in this study in Table~\ref{Tab:MESA_model_sets}. The main model set of this article is the model set of all possible and likely sdB progenitors accounting for the Galactic evolution of metallicity and star formation rates; Model set 1. This set consists of $2060$ MESA simulations based on a synthesised population described as follows.

First, we initialised a larger population of systems representing all the binaries in the Galaxy  (in practice, about $10^4$ times larger then the number of MESA simulations). We used the Kroupa-Haywood v6 initial mass function \citep{Kroupa2008, Haywood1997} as implemented in the Besançon Galactic model by \citet{Czekaj2014}. Therein, the IMF, ${\rm d}N/{\rm d}m\sim m^{-\alpha}$, is set by slope $\alpha$ equal to $1.3$ for $m$ between $0.09\,M_\odot$ and $0.5\,M_\odot$, to $1.8$ for $m$ between $0.5\,M_\odot$ and $1.53\,M_\odot$ and to $2.3$ for $m$ between $1.53\,M_\odot$ and $120\,M_\odot$. After drawing the mass of the primary, we drew its Galactic bin by using bin mass fractions in Table~\ref{Tab:GPop} as weights. Then we initialised the age by drawing uniformly in the age range of the Galactic bin and chose the metallicity by drawing a Gaussian variable with the parameters from Table~\ref{Tab:GPop}. We drew the companion mass assuming a uniform distribution in $M_{\rm comp}/M_{\rm primary}$ \citep{Raghavan2010} and the orbital period $P$ assuming a flat distribution in $\log P$ in the range between $1$ and $10^4$ days \citep{Abt1983}. We assumed circular binaries since red giants efficiently circularise before overflowing the Roche lobe \citep{Vos2015}.

For the actual MESA simulations, we selected from the large population described above only those binaries which may, in principle, result in long-period composite sdB binaries forming at present epoch. In particular, we considered only binaries with primary masses above $0.7\,M_\odot$, since lower-mass stars remain on the main sequence, and below $2.0\,M_\odot$, since higher-mass stars ignite the helium core non-degenerately \citep{Hurley2000} and are thus expected to lead to a different population of binaries. Furthermore, we selected the binaries with $M_{\rm primary}/M_{\rm comp}$ between $1$ and $3$ since higher mass ratios lead to unstable mass transfer on the giant branch, for example \citet{Tauris2000} and \citet{Pavlovskii2015}. We selected the period to be in the range between $0.1$ and $1.4$ of the period corresponding to Roche-lobe overflow at the tip of the red giant branch as determined from single evolutionary tracks from MIST database \citep{MIST}. The lower limit of $0.1$ corresponds to Roche lobe overflow on the sub-giant branch, at which point the helium cores do not have enough mass to ignite. The higher limit of $1.4$ was set to ensure that the binary MESA runs lead to Roche lobe overflow on the red giant branch in agreement with the single track-based estimates from the MIST database. Finally, we selected the binaries which become sdB stars at the present time. Specifically, we selected the binaries in which the primary reaches the tip of the red giant branch (based on single evolutionary tracks from MIST) not earlier than $300\,{\rm Myr}$ before the present day and not later than $700\,{\rm Myr}$ after the present day. If the tip was reached earlier than $300\,{\rm Myr}$ before the current time then, even allowing for possible $100\,{\rm Myr}$ difference between MIST and actual MESA tracks, the sdB star would have already reached the end of its lifetime and not be observed as an sdB star at present. If the tip was reached more than $700\,{\rm Myr}$ after present, then even allowing the possible difference of $100\,{\rm Myr}$ between MIST and MESA, the primary would still be on the main sequence at the present time and not able to produce an sdB. This way, through the selection process, the initial sample of $3.0\cdot 10^7$ general Galactic binaries was reduced to a sample of $2060$ systems which may potentially produce sdB binaries today. By examining the results of our MESA simulations, we ensured that all the produced sdB binaries originated from the systems fully enclosed within the parameter space we selected here.

A large fraction of the systems in the main set of runs (Set 1) are members of the thin disc as it is the most massive component of the Galaxy, whereas the thick disc and the halo, correspondingly, contribute little or do not contribute at all to the systems in the figure. To investigate the $P\,-\,q$ relation caused by the different components of the Galaxy in more detail (specifically, the thick disc, halo and bulge), we created separate model sets focused on those populations. These additional sets are numbered 1.1, 1.2 and 1.3 for, respectively, the thick disc, bulge and the halo and are summarised in Table~\ref{Tab:MESA_model_sets}.

To estimate the effect of using the Galactic model compared to more simple metallicity prescriptions, we simulated $1200$ binaries at fixed metallicities of $0$, $-1$, and uniformly distributed metallicities between $-1$ and $0.25$, also focusing on the sdB systems forming in the Galaxy today and keeping all the other parameters the same. These runs correspond to runs 2, 3 and 4 in Table\,\ref{Tab:MESA_model_sets}.

To demonstrate the importance of using the MESA code for producing sdB binaries as opposed to using synthetic codes which do not evolve stellar structure, we have re-simulated the main set of runs with the BSE code \citep{Hurley2002}, which corresponds to Set 5 in Table\,\ref{Tab:MESA_model_sets}. To reproduce the evolution described in Section~\ref{sec:BinEvol}, we also had to modify the BSE code. The default implementation of red giant mass transfer in BSE significantly underestimates the critical initial mass ratio leading to stable mass transfer, for example \citet{Tauris2000} and \citet{Pavlovskii2015}. Therefore, we replaced the stability criterion based on simplified polytropic models by the stability condition $q_i=M_{\rm primary}/M_{\rm comp}\leq 2.0$, to represent the typical critical mass ratios we observe in MESA runs, as we present further in Section~\ref{sec:ResMain}. Furthermore, we have modified the code to produce mass loss with the angular momentum of the accretor. Finally, we have removed the limiter on the red giant mass transfer rate which is set to be the thermal-timescale mass transfer rate of about $10^{-6}\,M_\odot/{\rm yr}$, since in MESA runs red giants reach mass transfer rates well above that value.
 
Finally, to explore in detail the effects of different initial binary parameters on the final periods and mass ratios, we have simulated additional 121 binaries using the MESA code, which corresponds to Set 6 and which we present in detail in Section~\ref{sec:effect_initial_parameters}.

To obtain the Galactic formation rates from our study, we used the total stellar Galactic mass of $6.43\cdot 10^{10}\,M_\odot$ \citep{McMillan2011}, the overall binary fraction of $0.45$ and the close binary fraction (sub-fraction of the binaries which are in period range of $1$ to $10^4\,{\rm d}$) of $0.25$, suitable for the mass range and the typical metallicities considered here \citep{Moe2019}. With these parameters set, we could identify how many stars in the Galaxy produce one sdB-progenitor binary and this way, by considering the sdB formation rate in our simulations, to re-normalise it to the sdB formation rate in our Galaxy. We calculated the present-day number counts of different sub-groups of sdB binaries by using their mean harmonic lifetimes represented by their core-helium burning phase. We ignored the effects of anti-correlation of the close binary fraction and metallicity in this study \citep{Moe2019} to identify the effect of purely-Galactic metallicity evolution.

\begin{table}
\begin{center}
\caption{Summary of the set of runs performed in this study. The columns show the name of the model set (Set), the metallicity [Fe/H] and star formation rates (SFR) adopted in the runs, the total number of simulated systems in each set (Runs), the total number of sdBs produced (sdBs) and the number of sdBs that are observable as composite sdB binaries (Obs).}
\begin{tabular}{|c|l|l|c|c|c|}
 \hline
 Set & [Fe/H] & SFR & Runs & sdBs & Obs \\ 
 \hline
 1   & MW - Full                & MW    & 2060 & 149 & 89 \\
 ~   & \hspace{0.6cm} - Thin    & ~     & 1681 & 129 & 77 \\
 ~   & \hspace{0.6cm} - Think   & ~     & 153  & 11  & 8  \\
 ~   & \hspace{0.6cm} - Bulge   & ~     & 224  & 9   & 4  \\
 ~   & \hspace{0.6cm} - Halo    & ~     & 2    & 0   & 0  \\
 1.1 & MW - Thick only          & burst & 450  & 30  & 15 \\ 
 1.2 & MW - Bulge only          & burst & 360  & 27  & 19 \\ 
 1.3 & MW - Halo only           & burst & 450  & 23  & 8  \\ 
 2   & Fixed at -1              & const & 360  & 49  & 27 \\
 3   & Fixed at 0               & const & 360  & 47  & 16 \\
 4   & -1 to 0.25               & const & 360  & 35  & 14 \\
 5   & MW - Full                & MW    & 2060 & 55  & --- \\
 ~   & \hspace{0.6cm} - Thin    & ~     & 1681 & 44  & --- \\
 ~   & \hspace{0.6cm} - Thick   & ~     & 153  & 5   & --- \\
 ~   & \hspace{0.6cm} - Bulge   & ~     & 224  & 6   & --- \\
 ~   & \hspace{0.6cm} - Halo    & ~     & 2    & 0   & --- \\ 
 6   & Synthetic, 1-$\sigma$    & synth. & 121  & 91  & 91 \\
 \hline
\end{tabular}
\tablefoot{All sets of runs were performed with MESA code, with the exception of Set 5, which was performed with BSE code. For the models calculated with the BSE code, it is not possible to determine if they are observable.}
\label{Tab:MESA_model_sets}
\end{center}
\end{table}

\section{Results} \label{sec:ResMain}

Our main set of 2060 runs (model set 1) has produced $149$ sdB and $40$ sdA binaries, of which $89$ sdB and $28$ sdA binaries would have been identified as long-period composite subdwarf binaries in actual observations following the criteria described in section \ref{sec:ObsCond}.

\subsection{$P\,-\,q$ relation}


\begin{figure*}
    \includegraphics{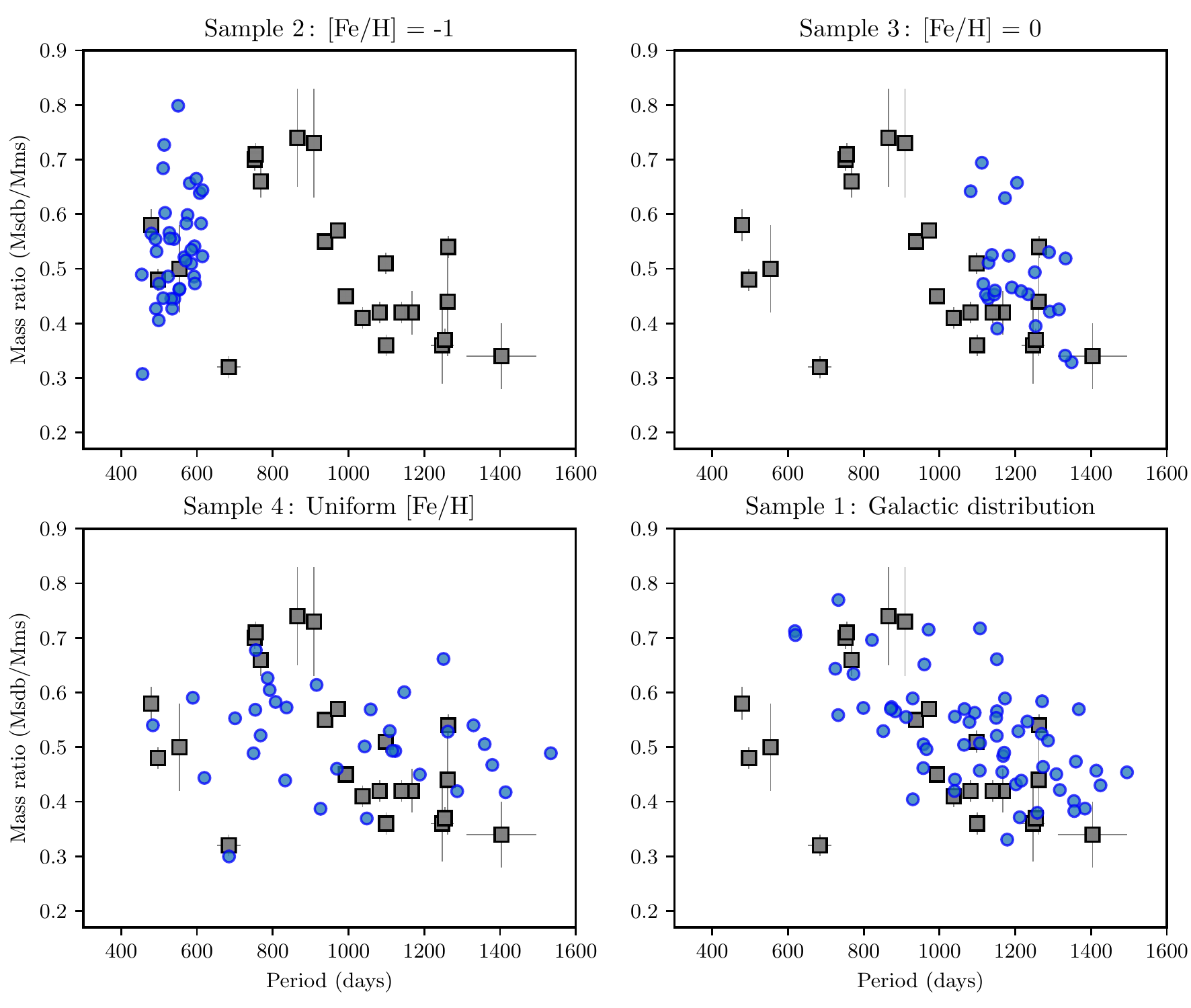}
    \caption{The period-mass ratio relation of the sdB binaries produced by the different MESA samples. The grey squares show the real observed systems, while the blue circles show the MESA models which would have been observationally identified as composite long-period sdB binaries, i.e. fulfil the visibility criteria in Section~\ref{sec:ObsCond}. The observed systems, located within the nearest $1\,\textrm{kpc}$, belong to the solar neighbourhood. Most of the simulated sdB binaries in the Galactic population belong to the thin disc, which is the dominant Galactic component (see Table~\ref{Tab:MESA_model_sets}).}
    \label{fig:metalicity_P_q_comparison}
\end{figure*}

Our models can explain the main branch of the $P\,-\,q$ relation in long-period sdB binaries by using a simple Galactic chemical evolution model and without any free parameters, which is the most important result of this study. In Fig.\,\ref{fig:metalicity_P_q_comparison}, we show the  $P\,-\,q$ relation of the models obtained for several different metallicity distributions from Tab.~\ref{Tab:MESA_model_sets}. In these plots, the blue circles show the produced systems which would have been observed as hot composite sdB binaries, that is fulfil the visibility criteria described in Section~\ref{sec:ObsCond}. The real observations are shown in grey rectangles.

The model set with fixed metallicity and constant star-forming rate (model sets 2 and 3) can only cover a small part of the observed $P\,-\,q$ relation. This result is expected as the RGB radius is strongly dependent on the metallicity, and as the system needs to interact when the RGB core is close to the He ignition mass, the period range at which the interaction can take place and form an sdB is small. The sample with a sub-solar metallicity of ${\rm [Fe/H]} = -1.0$ has orbital periods around $500$ days while the solar metallicity sample reaches orbital periods around $1200$ days. The final mass ratio range is similar for both samples and is between $q = M_{\rm sdB}/M_{\rm comp} = 0.3$ and $0.8$. This result is consistent with the idea that the main effect of metallicity is on the final periods rather than mass ratios. We demonstrate this effect in more detail in Section~\ref{sec:effect_initial_parameters}. The sample with a sub-solar metallicity of $-1.0$ covers most of the systems in the secondary branch of the $P\,-\,q$ relation.

The model set with a uniform metallicity distribution (model set 4) can cover all observed orbital periods, including those of the secondary branch. However, this sample can not, in any meaningful way, explain the observed $P\,-\,q$ correlation of the main group. As the metallicity can freely vary between ${\rm [Fe/H]} = -1.0$ and $0.25$, sdB binaries with orbital periods between $\sim 500$ and $\sim 1400\,{\rm days}$ can be created, as expected from model sets 2 and 3.

\begin{figure}
    \includegraphics{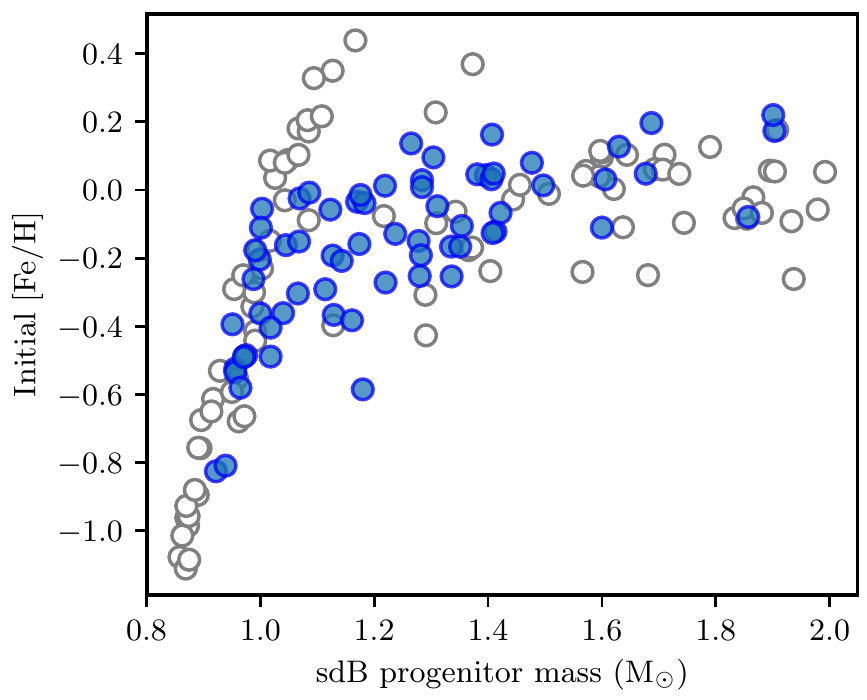}
    \caption{The initial metallicity as a function of the sdB progenitor mass for all systems in the model sample following the Galactic model (sample 4). Systems that fulfil the criteria to be recognizable as a composite sdB binary are plotted in filled blue circles, while those where only one component is visible are shown in open grey circles.}
    \label{fig:Galactic_FeH_Minit}
\end{figure}

The last model set in Fig.\,\ref{fig:metalicity_P_q_comparison}, which follows the Galactic model (model set 1), covers the $P\,-\,q$ correlation in the main group very well, predicting systems with high mass ratios at short orbital periods and low mass ratios at long orbital periods. The relation causing this pattern is shown in Fig.\,\ref{fig:Galactic_FeH_Minit}. This figure shows the sdB progenitor mass as a function of the initial metallicity of the system. The systems with a lower-mass sdB progenitor formed when the Galaxy was younger and consequently have a lower metallicity. These systems need to have a shorter orbital period to interact when the sdB progenitor has a sufficiently massive core and will produce shorter orbital period systems. Since lower-mass progenitors also have lower-mass companions, the resulting sdB binaries have high final mass ratios $q=M_{\rm sdB}/M_{\rm comp}$ and end up on the top of the $P\,-\,q$ relation. Higher-mass massive progenitors, on the opposite, have higher metallicities, hence higher final periods, and higher-mass companions, and hence lower final mass ratios. As a result, higher-mass progenitors lead to sdB binaries at the bottom of the $P\,-\,q$ relation. For a given sdB progenitor mass and orbital period, there is only a small range in mass ratios and initial periods that will be stable but still have sufficient mass loss and helium core size to result in an sdB star. We further explore the effects of all the initial parameters in detail in Section~\ref{sec:effect_initial_parameters}.

\begin{figure}
    \includegraphics{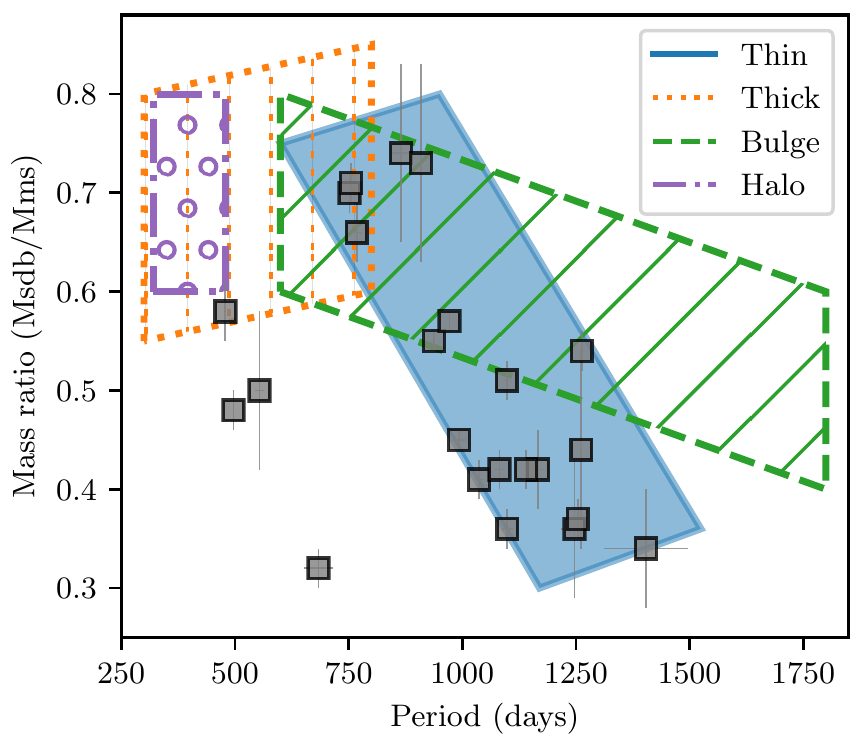}
    \caption{Schematic plot showing where wide sdBs formed from the different Galactic sub-populations are located in the period -- mass ratio diagram. Thin-disc objects are shown in blue, thick-disc objects are shown in orange, bulge objects are shown in green and halo objects in purple. The observed systems are plotted in grey squares.}
    \label{fig:P_q_populations}
\end{figure}

The different Galactic populations outlined in Table\,\ref{Tab:GPop} can populate different parts of the $P\,-\,q$ diagram. This is shown in Fig.\,\ref{fig:P_q_populations} for the four main Galaxy components: the thin disc, the thick disc, the halo and the bulge. Since the thin disc is the most massive component of the Galaxy, most modelled systems are located in the thin disc. The Galactic bulge, which is an old high-metallicity population, will produce sdB binaries at long orbital periods, to the right of the currently observed $P\,-\,q$ relation, as well as sdB binaries which overlap with the currently observed $P\,-\,q$ relation. Systems from the thick disc have a lower metallicity and can be found at slightly shorter orbital periods. The halo contributes so little to the Galactic mass (Tab.\,\ref{Tab:GPop}) that there are no sdB binaries, observable as composite systems, formed from the halo population in the main sample. As follows from the halo set of runs (Set 1.3), members of the halo population have short periods below $500\,{\rm d}$ and are located to the left of the thick disc population. While no Galactic population in our model covers the systems in the secondary branch in the $P\,-\,q$ relation, Fig.\ref{fig:metalicity_P_q_comparison}, we discuss the possible origins for these objects in Section~\ref{sec:DiscStellar}.

We have also verified if synthetic stellar population codes, such as BSE, can reproduce the $P\,-\,q$ relation of sdB binaries. By running the same Galactic population as in model set 1 through the BSE code (model set 5), we found that even when modified to have the same angular momentum-loss prescription and similar stability criteria as MESA, all the sdB-binaries acquired orbital periods below $400\,{\rm days}$. This result agrees with the original models by \citet{Han2003}. As also discussed in \citet{Chen2013}, these binaries fall into a region of the  $P\,-\,q$ plane which is quite far away from the observed systems. Our study shows that the reason why the BSE code does not reproduce the $P\,-\,q$ relation is not the angular momentum loss prescription. Indeed, in contrast to our MESA results, even with the same angular momentum loss prescription as in MESA, no binaries with initial periods longer than $300\,{\rm days}$ have produced sdB objects in the BSE sample. In other words, in BSE, sdB binaries were produced by progenitors with incorrectly-short periods. The too short periods of these progenitors explain the too short periods of the resulting sdBs. In contrast, the objects which would have been progenitors in MESA did not produce any sdB binaries in BSE. Therefore, we conclude that the reason why BSE produces too short periods of sdBs is the incorrect criteria by which the BSE code defines the formation of stripped helium stars. The BSE code is based on interpolating over the models of single stars, whereas the formation of sdB-binaries is strongly sensitive to the details of binary mass transfer, for example \citet{Chen2013}. Therefore, the conditions, by which the BSE code defines whether a stripped He star has formed, are not consistent with the more detailed MESA simulations.

\subsection{$M_{\rm sdB}$ -- Metallicity -- $P_{\rm final}$ correlations}

\begin{figure*}
    \includegraphics{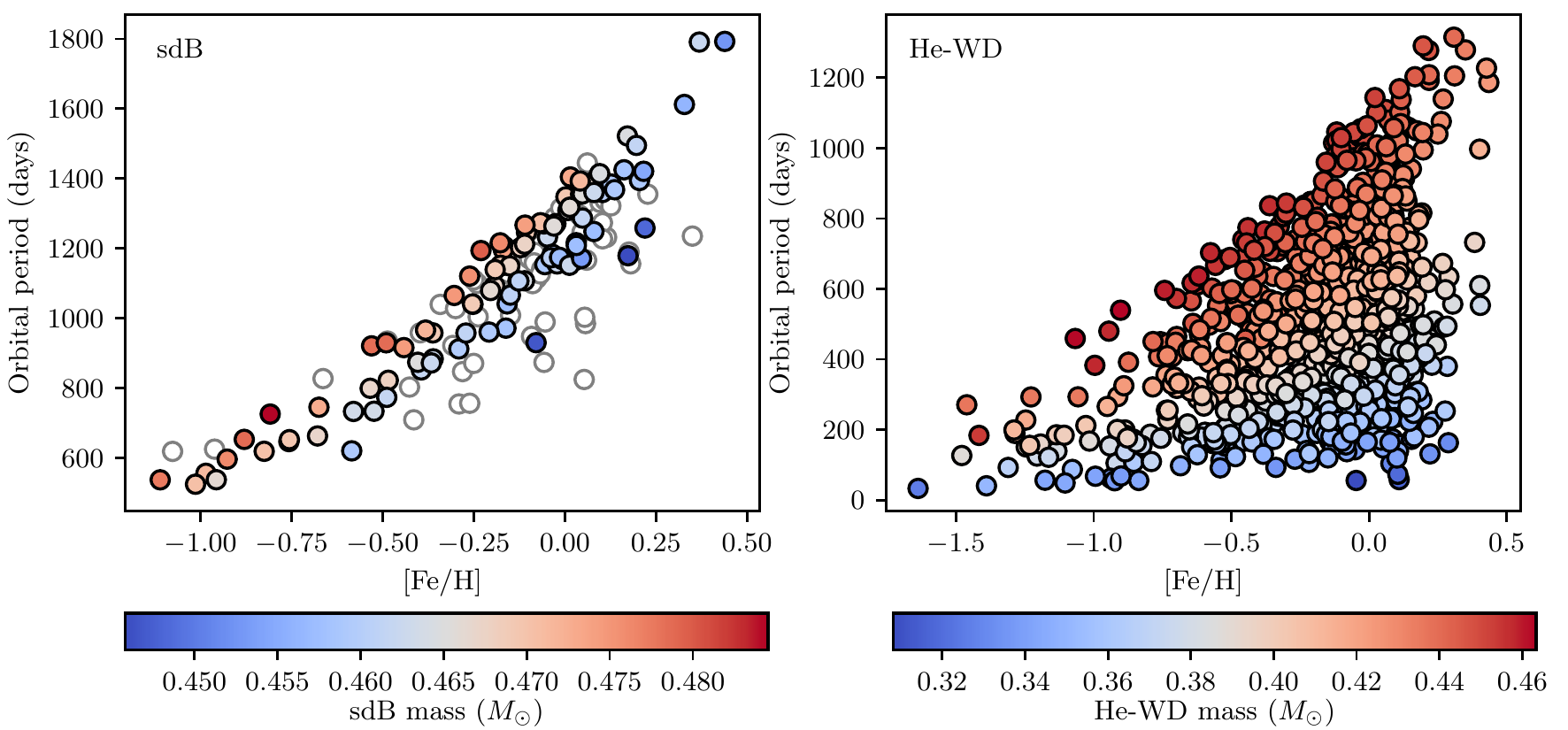}
    \caption{The relation between the orbital period and the metallicity for the sdB binaries (left) and He-WD binaries (right) in our sample. The relation is a result of the donor radius on the RGB being strongly sensitive to its metallicity. The period-metallicity relation also depends on the sdB/He-WD mass, which is shown by the colour scale. The systems that are recognizable as composite sdBs are shown in filled circles while the single-lined ones are shown in grey open circles. For the He-WDs no distinction is made between composite and single-lined systems.}
    \label{fig:Pfinal_FeH_Mfinal}
\end{figure*}

Since the maximum red giant radius strongly depends on metallicity, the final orbital periods of sdB binaries must also correlate with their initial metallicity. We show this relation in Fig.\,\ref{fig:Pfinal_FeH_Mfinal}. While the periods and metallicities have been known to correlate for hot subdwarfs, \citep[e.g.][]{Chen2013}, and other post-mass-transfer systems, such as low-mass and extremely-low-mass WDs, \citep[e.g.]{Istrate2016}, this relation has never been depicted for a MESA-based synthesised population which follows the Galactic metallicity and star formation history. Neither has it been compared to actual observations. 

The correlation of the final orbital periods with the initial metallicity may potentially be used to test the model. Moreover, the correlation is stronger for sdB binaries compared to other post-mass transfer systems, which is because, for a given metallicity, the red giant radius is the dominant parameter determining the final period (as we discuss in Section~\ref{sec:effect_initial_parameters}). Indeed, the range of possible sdB masses is very narrow (in all our models, between $0.44$ and $0.49\,M_{\odot}$) and the range of initial periods and mass ratios leading to sdB binaries is also fairly narrow (at most, $40\,\%$). Therefore, the relation between the period and initial metallicity of sdB binaries is very tight and has a very small spread, as shown in the left panel of Fig.\,\ref{fig:Pfinal_FeH_Mfinal}.

\begin{figure}
    \includegraphics{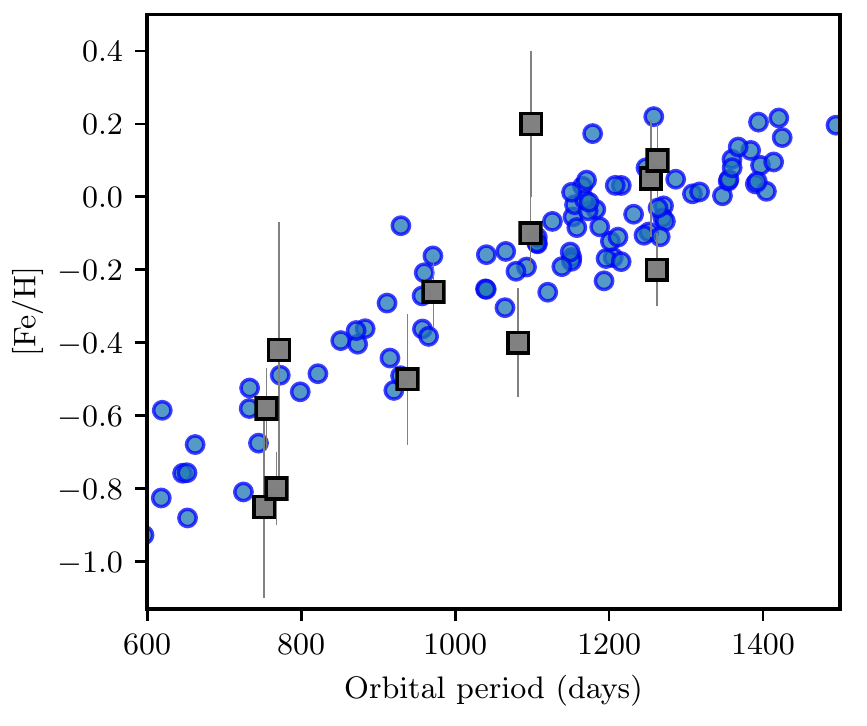}
    \caption{The observed versus modelled orbital period and metallicity correlation for long-period sdB binaries. The black squares show all the observed wide sdB binaries for which such measurements are available. The results from our MESA model following the Galactic model (model set 4) are shown in blue circles. Only model systems that fulfil the observability criteria are shown.}
    \label{fig:observed_period_feh}
\end{figure}

Only 12 wide composite sdB binaries have measurements of their metallicities available in the literature \citep{Vos2017, Vos2018, Vos2019}. These measurements have, on average, large errors as the spectra used to derive the metallicities have low signal-to-noise ratios. In Fig.\,\ref{fig:observed_period_feh}, the observed period-metallicity relation is shown together with the predictions from our models. The observations match very well with the predictions, and this way provide independent verification of our model.

For comparison, we show the same relation for the He WD outcomes in our Galactic sample in the right panel of Fig.\,\ref{fig:Pfinal_FeH_Mfinal}. These are systems in which the donor star lost too much mass to ignite He, and which end as He WDs. They range in mass between $0.30$ and $0.46\,M_{\odot}$. The He WDs show the same relation of longer orbital periods at higher metallicity for systems with the same final donor mass. However, as the mass range for He WDs and their companions is much larger than for sdBs, and as the mass of a He WD is not easy to determine accurately, there is no simple relation that can be tested from the observations of He WDs.

The mass of an sdB star formed from a low-mass progenitor igniting He under degenerate conditions is close to the canonical mass of $0.47\,M_{\odot}$ but varies with orbital period and initial metallicity. The MESA models here can be used to derive a fitting equation of the sdB mass as a function of the orbital period and metallicity. An equation that is linear in these two parameters can reproduce the sdB mass with an error less than 0.1\,\%:
\begin{equation}
    M_{\rm sdB} = 0.395 + 0.005 \cdot \frac{P_{\rm orb}}{100\,{\rm d}} - 0.058 \cdot {\rm [Fe/H]} \label{eq:Msdb_P_feh}
\end{equation}
A similar equation with an error less than 0.2\,\% can be derived for the He-WDs in our sample:
\begin{equation}
    M_{\rm WD} = 0.335 + 0.012 \cdot \frac{P_{\rm orb}}{100\,{\rm d}} - 0.048 \cdot {\rm [Fe/H]} \label{eq:Mwd_P_feh}
\end{equation}
It is important to notice that equation~\ref{eq:Mwd_P_feh} applies to the Galactic binaries in which the WD has formed recently (within the last few $100\,{\rm Myr}$) and is only valid for the parameters range covered in our models. This range corresponds to the same range that is shown in Fig.\,\ref{fig:Pfinal_FeH_Mfinal}.

\subsection{Linking to the progenitor properties}

\begin{figure}
    \includegraphics{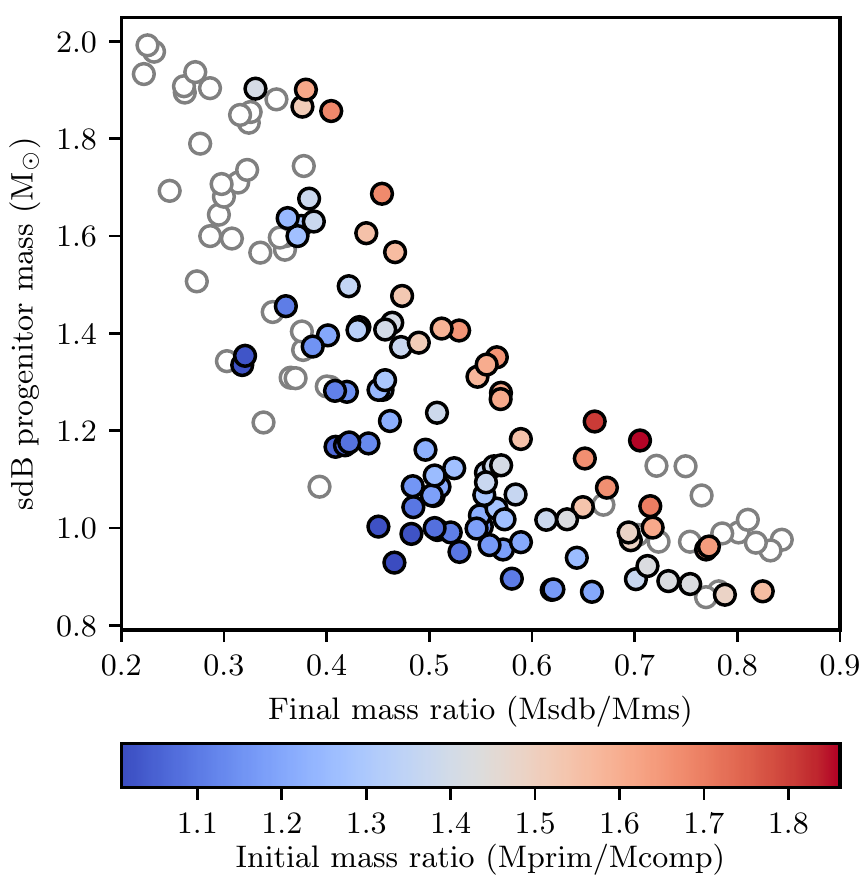}
    \caption{The correlation between the mass ratios in the observable composite sdB binaries (filled circles) and the initial sdB progenitor masses in our models. The colour shows the initial mass ratios and the empty circles show the sdB binaries, which would not be observationally identified as composite systems. One may see that the observed mass ratios constrain, to a certain extent, the initial progenitor masses of sdB binaries.}
    \label{fig:qfinal_M1init}
\end{figure}
\begin{figure}
    \includegraphics{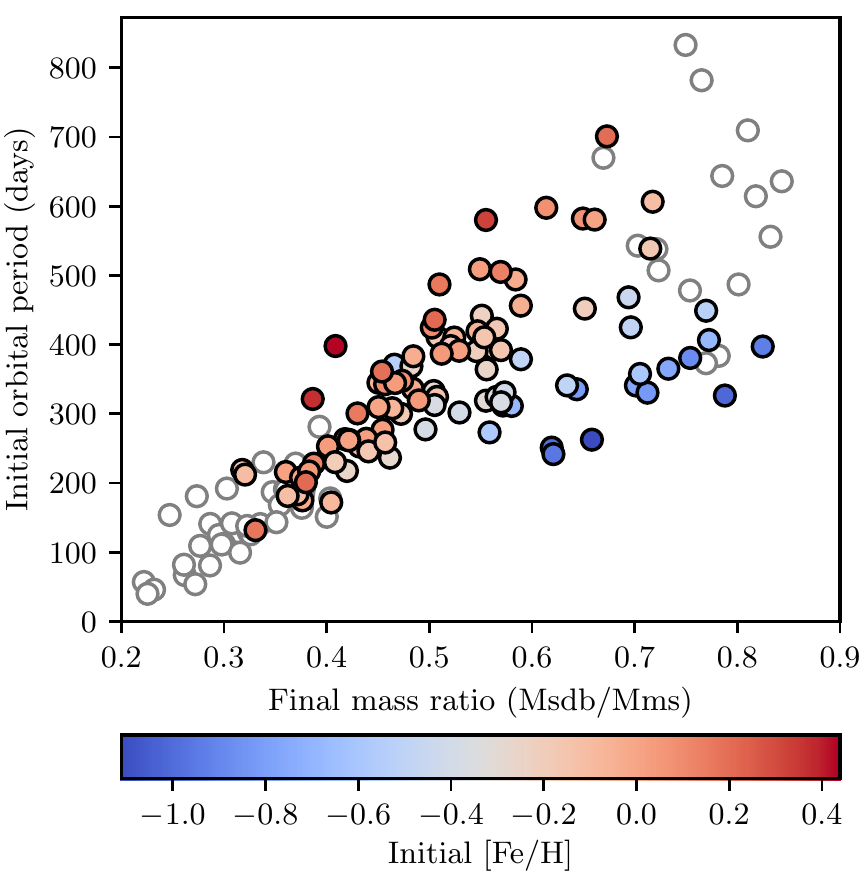}
    \caption{The correlation between the mass ratios in the observable composite sdB binaries (filled circles) and the initial orbital periods of the systems in our models. The colour shows the initial metallicity and the empty circles show the sdB binaries, which would not be observationally identified as composite systems. Given the rather tight correlation, one may constrain the initial period and, in certain cases, the progenitor metallicity from the present-day mass ratios.}
    \label{fig:qfinal_Pinit}
\end{figure}

Our models offer an interesting possibility of connecting the properties of the observed sdB systems to their progenitors. Such an inference is possible because several present-day observables are strongly correlated to the initial parameters of the binaries. The two strongest correlations are between the final mass ratio and the sdB progenitor mass, and between the final mass ratio and the initial orbital period. These relations are shown in Fig.\,\ref{fig:qfinal_M1init} and \ref{fig:qfinal_Pinit}.  These correlations can be applied to the data from ongoing large spectroscopic surveys, such as APOGEE, Gaia-ESO, WEAVE and 4MOST. These surveys, together with Gaia DR3, will provide a large sample of wide main-sequence binaries with periods and metallicities, for example \citet{PriceWhelan2020} and \citet{Hayden2015}. These data sets can be used to verify our models further and to connect the sdB populations to their progenitors.

In the first figure, it may be seen that systems with a larger final mass ratio have a lower sdB progenitor mass. As may be seen from the figure, the spread in this relation is caused by the initial mass ratio. By examining the spread, we see that, for a given sdB progenitor mass, there is a limited range in companion masses that will lead to an sdB star. If the companion mass is outside of that range, a He-WD or a horizontal-branch (HB) star is formed instead.

The second strong correlation is between the initial orbital period and the final mass ratio. This correlation shows that systems with larger observed mass ratios are produced from systems with larger initial orbital periods. For a given observed mass ratio, the initial orbital period also varies with the metallicity of the system. When both the mass ratio and the metallicity can be determined, the initial orbital period can be derived quite precisely.

\begin{figure}
    \includegraphics{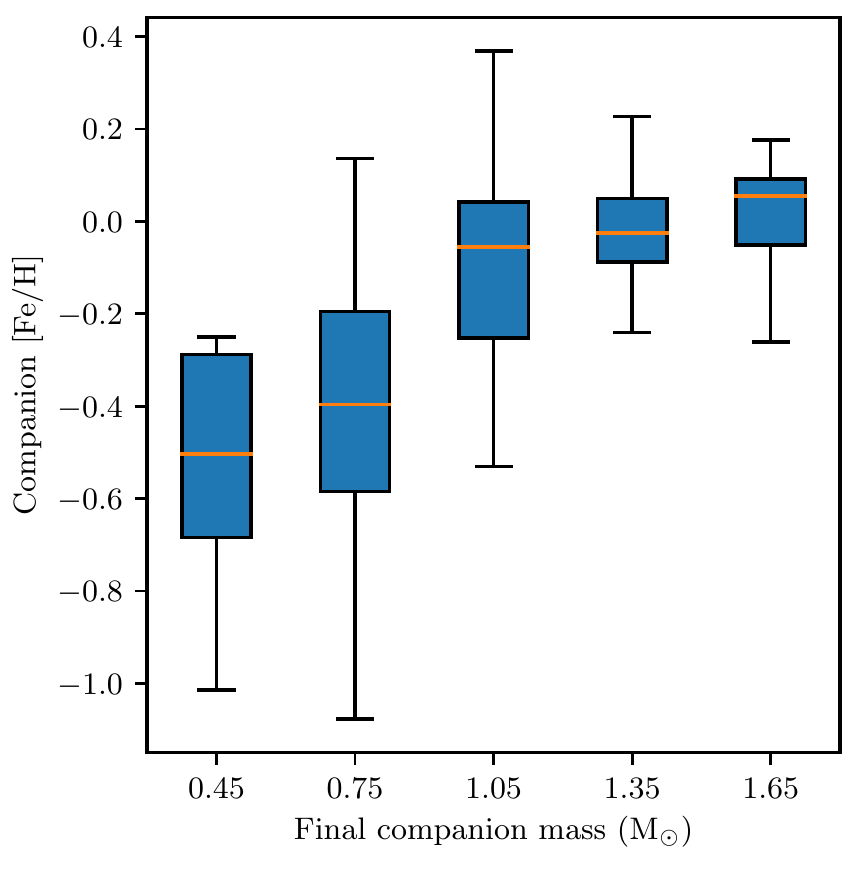}
    \caption{The metallicity of the companion stars as a function of the companion mass after at the binary sdB formation time. The box plot shows the average metallicity with an orange line, the 1-sigma range with a box and the 3-sigma range with whiskers. It may be seen that the present-day metallicity and companion masses of sdB binaries are correlated, according to our model.}
    \label{fig:M2final_FeHinit}
\end{figure}

Finally, as we show in Fig. \ref{fig:M2final_FeHinit}, our models imply that the companion masses in the observed sdB binaries should correlate with their metallicities. This correlation may, in principle, be used to constrain our model, or, alternatively, the metallicity history of the Galaxy. The advantage of this correlation is that these properties do not require the determination of the orbital parameters, which makes it easier to obtain a large sample in less time.

\subsection{Sensitivity to initial parameters}\label{sec:effect_initial_parameters}

\begin{figure*}
    \includegraphics{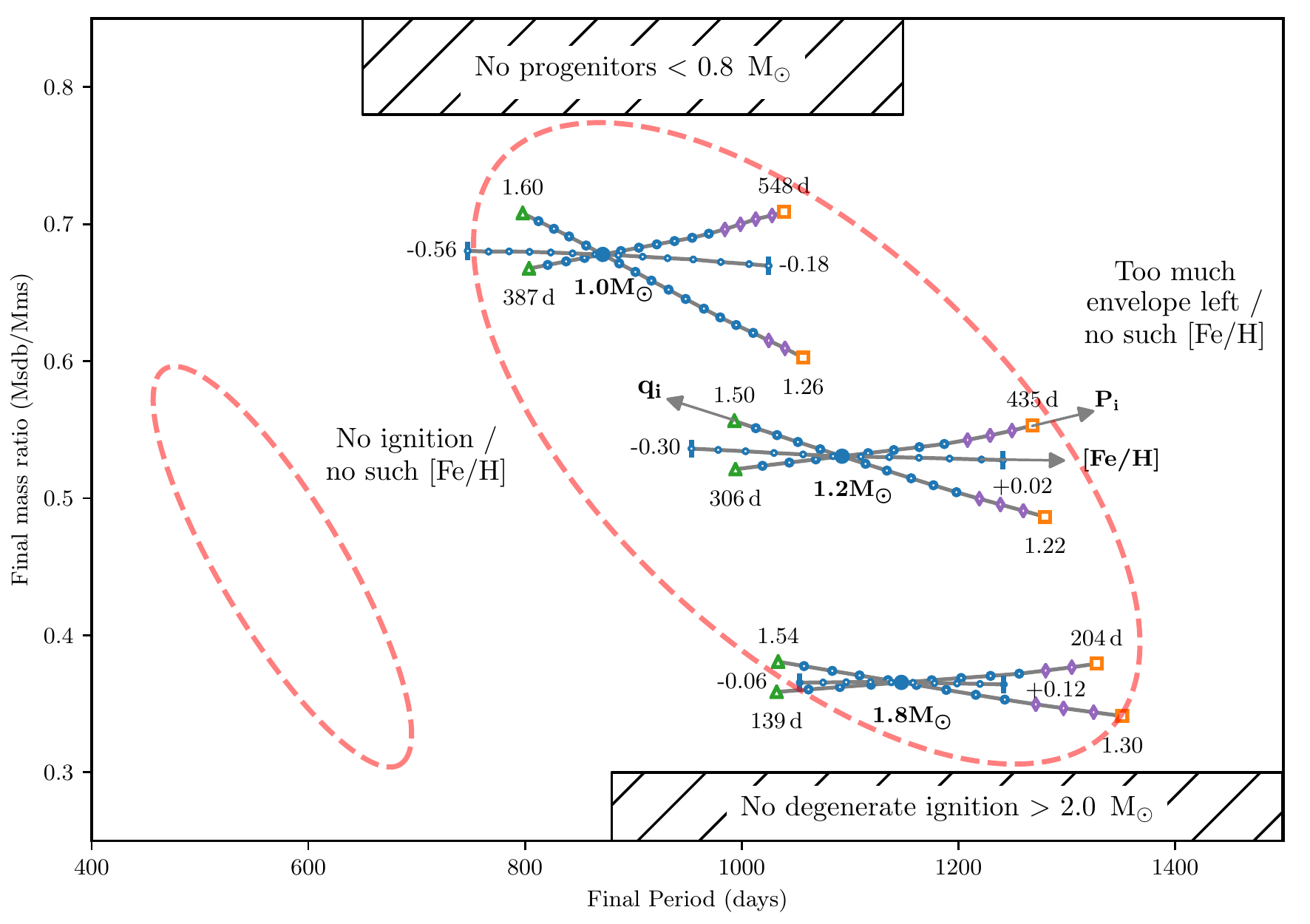}
    \caption{This figure shows the effect of the initial parameters: orbital period (P$_{\rm i}$), mass ratio (q$_{\rm i}$), metallicity ([Fe/H]) and sdB progenitor mass on the final orbital periods and mass ratios of sdB binaries. We consider primary masses of $1.0$, $1.2$ and $1.8\,M_{\odot}$, indicated in boldface below each set of tracks. Systems that result in hot composite sdB stars are marked with blue circles, composite sdA stars are marked with purple diamonds, while systems resulting in He-WDs or horizontal branch stars are marked with green triangles and orange squares respectively. The locations of the main and second branches of the observed $P\,-\,q$ relation are indicated with the red dashed ellipses. The initial mass ratios and orbital periods are varied in equal steps in the range that still produces hot composite subdwarf stars for the given initial donor mass, while, for the metallicity, we show the one-sigma range of the Galactic metallicity corresponding to the time when the progenitors of each given mass formed (Tab.~\ref{Tab:GPop}). This figure clearly shows that for a given initial donor mass, only a small range in the final mass ratios can be reached by varying the other initial parameters. Varying the metallicity has almost no effect on the final mass ratio, but strongly changes the final orbital period. The correlation between the final mass ratio and the initial donor mass combined with the link between metallicity and initial donor mass due to Galactic evolution then explains the observed $P\,-\,q$ relation. See section \ref{sec:effect_initial_parameters} for details.}
    \label{fig:effect_initial_parameters}
\end{figure*}

To investigate the sensitivity of the final parameters to the initial parameters, we ran a dedicated set of models by varying each initial parameter separately (Set~6 in Tab.~\ref{Tab:MESA_model_sets}). The results of these runs are shown in Fig.\,\ref{fig:effect_initial_parameters}. We chose models with three representative initial donor masses of $1.0$, $1.2$ and $1.8\,M_{\odot}$, each of which leads to a long-period hot composite sdB binary. For each of these three masses, we varied the initial mass ratio, metallicity and initial orbital, in equal steps, in the range that would still produce a hot subdwarf star after the mass-loss phase. If the initial mass ratio is too high, too much mass is lost and the donor ends up as a He-WD. If the mass ratio is too low, too little mass is lost, and the donor ends up as a horizontal branch star. Similarly, if the initial period is too short, the red giant core at the end of mass transfer is too small in mass to ignite helium, whereas if the period is too large, too much envelope remains at the end of mass transfer to produce an sdB star. The range in metallicity that can produce a hot subdwarf star is large. However, due to Galactic evolution, only a small range in metallicity is possible for each initial donor mass. In Fig.\,\ref{fig:effect_initial_parameters} we, therefore, show only the one-sigma range of possible metallicities for each initial donor mass.

From these models, it is clear that the dominant factor determining the final mass ratio is the initial donor mass. This dependency may be understood as follows. Lighter donors have on average lower-mass companions, whereas heavier donors have on average higher-mass companions. In contrast, the final sdB masses only weakly depend on the primary mass. Similarly, since companions accrete only small amounts of mass, as we discuss in Section~\ref{sec:BinEvol}, their final masses remains close to their initial masses. Therefore, lighter donors with lower-mass companions lead to larger final mass ratios $q=M_{\rm sdB}/M_{\rm comp}$, and heavier donors with higher-mass companions lead to lower final mass ratios. This relation between the initial donor mass and the final mass ratios is also shown in Fig.\,\ref{fig:qfinal_M1init}.

For a given initial donor mass, the final mass ratio will only change little by varying the initial mass ratio or the initial orbital period. For example, for the system with an initial donor mass of $1.2\,M_{\odot}$, the final mass ratio can vary from $0.48$ to $0.55$ by changing the initial mass ratio or the orbital period. Similarly, changing the metallicity of the system only has a minimal effect on the final mass ratio. It is also important to notice that systems with a lower initial donor mass have a higher final mass ratio and vice versa. 

The initial metallicity affects the final orbital period significantly. The one-sigma range in metallicity can change the final orbital period by $200$ -- $300$ days. Since heavier donors have higher metallicities and thus create sdB binaries with longer final orbital periods than the lighter donors, there is a correlation between the initial donor mass and final orbital period. Combining this and the earlier correlation between the primary mass and the final mass ratio then explains the observed period-mass ratio relation.

Long-period sdA binaries behave similarly to sdB binaries. The main difference between both types of systems is that sdA binaries have more envelope mass left and are therefore cooler (see Fig.\,\ref{fig:teff_logg_Hecoreburning}). In Fig.\,\ref{fig:effect_initial_parameters}, the sdA binaries are located on the right ends of the hot subdwarf series, just before they become horizontal branch stars. Therefore, the relation between orbital period and mass ratio that exists for sdB binaries also exists for sdA binaries. As these systems have more envelope mass left, their $P\,-\,q$ relation is shifted to longer final orbital periods.

\subsection{Mass distribution}

\begin{figure}
    \includegraphics{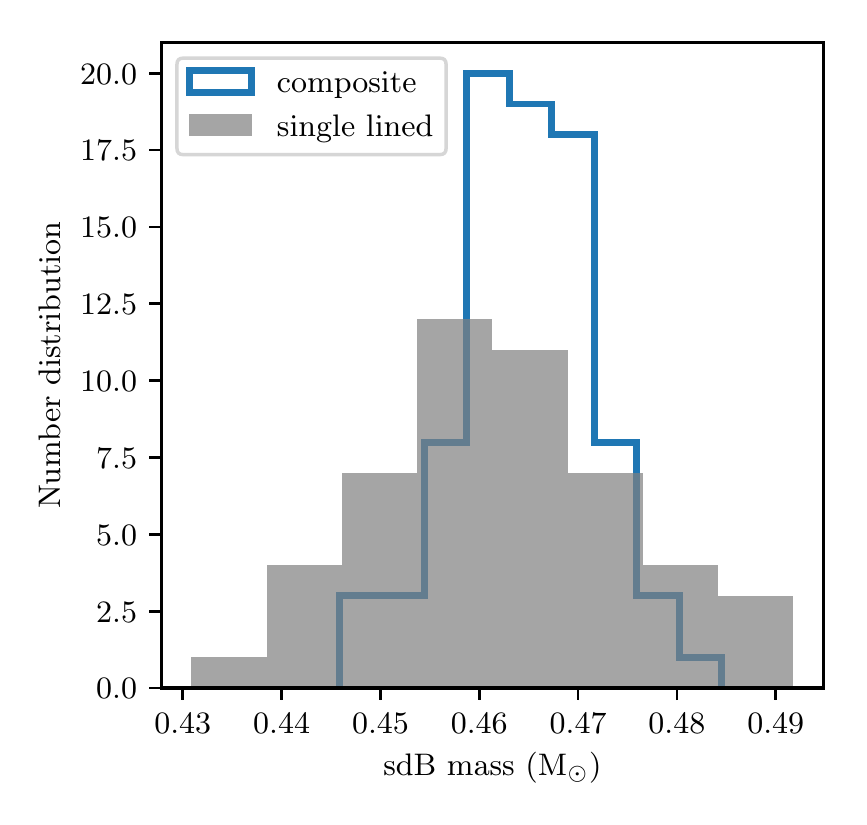}
    \caption{The mass distribution of the sdB stars in the Galactic sample. The distribution of the composite systems is shown with the blue line, while that of sdBs in single-lined systems is shown in grey. The bins were selected using Knuths rule \citep{Knuth2006}. Both distributions are similar and consistent with the empirically determined masses, for example \citet{Fontaine2012, Schneider2020}.}
    \label{fig:sdB_mass_distribution}
\end{figure}

The mass distribution of the sdB components in the Galactic sample is shown in Fig.\,\ref{fig:sdB_mass_distribution}. Mass distributions of the sdB stars in composite binaries and single-lined systems are shown, respectively, in blue and grey. Both distributions peak at similar masses, and have a similar spread. The mass distribution of the sdBs in composite systems peaks at $0.465\, M_{\odot}$ and has a spread of about $0.01\,M_{\odot}$. These results are consistent with the earlier results of \citet{Han2002}, and the observed distributions of for example \citet{Fontaine2012} and \citet{Schneider2020}.

\section{Discussion}
\label{sec:DiscMain}

We have shown that the main branch of the observed $P\,-\,q$ relation in long-period composite sdB binaries may be fully explained, without free parameters, through a standard binary interaction prescription implemented in the MESA code and a standard model for the chemical history of the Galaxy. In this section, we show that both the MESA code with the binary interaction model and the model for the Galactic evolution have been essential ingredients in our model. Our method can likely improve the modelling of other post-mass transfer binaries containing hot subdwarf and HB stars, white dwarfs and, possibly, compact objects. Furthermore, by applying the criteria for the observational selection of long-period composite sdB binaries, we have shown that our model predicts several observational correlations. These correlations may be used to probe the progenitor properties of sdB binaries, as well as to test the model. Finally, our study offers an exciting possibility of mutually constraining the models of the Galactic chemical evolution and the observed intrinsic properties of sdB binaries, as well as of other exotic binaries in the field.

\subsection{The importance of Galactic evolution}


It is well known that metallicity has an important effect on the evolution of stars. For binary systems, the initial metallicity determines, to a large extent, the red giant radius of the stars and thus dictates when the interaction phase starts on the RGB. Therefore, metallicity has an important effect on the orbital period distribution of post-interaction systems. However, metallicity is often ignored in binary population synthesis studies. Typical binary population studies use a fixed (usually solar) or, in the best case, a uniform metallicity distribution \citep[see e.g.][]{Izzard2009, Hamers2013, Wijnen2015, Stanway2018}.

Using a fixed or uniform metallicity distribution has a benefit of simplifying the models and has its merits in some cases. In a realistic setting, however, stellar metallicity is coupled to the stellar age due to Galactic evolution. This coupling is particularly important when dealing with low-mass stars that have long MS lifetimes. If such stars reach the RGB, they must have formed in sub-solar metallicity environments,  and are thus sensitive to metallicity changes in the history of the Galaxy. In our models, stars with masses smaller than about $1.4 - 1.6\,M_{\odot}$ are most sensitive to the metallicity increase with time, see for example Fig.\,\ref{fig:Galactic_FeH_Minit}. Similar sensitivity will likely manifest itself in properties of other Galactic populations which are experiencing or have recently experienced mass transfer on the RGB from a low-mass donor. These include short-period sdB binaries \citep{Heber2016}, symbiotic binaries \citep{Munari2019}, extremely low-mass WD binaries 
\citep{Pelisoli2019} and binary pulsars with a He-WD companion \citep{Istrate2014}, among others.

The importance of the cosmological metallicity history has been recently recognised in population studies of binary compact objects, such as black holes or neutron stars, due to its effect on massive stellar evolution, for example \citet{Lamberts2018}, \citet{Boco2019}, \citet{Chruslinska2019}, \citet{Neijssel2019} and \citet{Toffano2019}. At the same time, only very few Milky Way studies have considered a realistic metallicity history so far, for example \citet{Lamberts2019} and \citet{Olejak2020}, focusing on gravitational wave sources. We expect that using simple, realistic Galactic metallicity models, as adopted in this study, will be common in modelling other Galactic populations in the near future.

\subsection{The importance of using MESA for modelling sdB binaries}


Modelling sdB binaries requires the use of a binary evolution code such as MESA, rather than traditional binary population synthesis codes. Such codes are based on fits to evolutionary tracks of single stars in isolation and are not designed to model a realistic response of red giants to mass loss. In comparison, binary evolution codes continuously solve the stellar structure equations for both stars, this way accurately evolving red giants through the envelope loss and the subsequent He-flash phases which lead to the formation of sdB stars. Furthermore, population synthesis codes make use of simplified analytic prescriptions when determining outcomes of binary mass transfer, this way limiting, potentially, the correctness of the results. For example, the BSE code determines the stability of red giant mass transfer by using polytropic models of red giants. In its standard implementation, BSE predicts that most of the sdB models calculated in this article would have been unstable and would have entered a common-envelope phase.

In section \ref{sec:ResMain}, we showed that even when modified to reproduce similar stability criteria as observed in the MESA code and to follow the same binary interaction model, BSE code still fails to produce sdB binaries at the observed orbital periods. This issue is related to the incorrect initialisation of the He star stage in the BSE code compared to MESA. The reason for the difference may be because the BSE fits for initialising He stars were designed under the old assumption of polytropic models, which  lead to incorrect stability criteria for red giant mass transfer. Nevertheless, with the above caveats in mind, traditional BPS codes may certainly benefit significantly from using Galactic metallicity distributions.

We expect that binary evolution codes such as MESA will be increasingly more important in the future. While such codes are more computationally expensive than population synthesis codes (tens of hours versus milliseconds to model a binary on a single modern CPU), population samples of a few thousand objects can be efficiently studied with binary evolution codes. If the initial parameters are chosen carefully, a sample of a few thousand objects can be statistically significant. As the computational power increases over time, binary evolution codes will allow for modelling progressively more rare populations and also provide useful calibrations to synthetic codes.

\subsection{Observational tests and predictions}


The models in this article match the main $P\,-\,q$ branch of the observed wide sdB binaries very well. The model sets that use the same binary interaction model but with a simplified metallicity distribution (fixed or uniform [Fe/H]) can match the observed range of mass ratios, and in the case of uniform [Fe/H], also the range of orbital periods. However, none of these models can explain the correlation between the orbital period and mass ratio. It is now clear that this correlation is mainly an effect of the metallicity evolution in the Galaxy.

Our models also predict a very strong correlation between the orbital period, sdB core mass and metallicity. This correlation is due to the RGB radius dependence on metallicity, which is one of its main determining parameters. Of these quantities, the orbital period and metallicity can be observed directly. Even though only a few systems have a published value for the metallicity, and usually with a large error, the observations match the predicted period-metallicity relation very well (see Fig.\,\ref{fig:observed_period_feh}). This is the first time the observed period-metallicity correlation of long-period sdB binaries has been quantitatively reproduced by a population synthesis study.

We note that although the correlation of the sdB core mass and metallicity has been discussed in \citet{Chen2013}, the period range obtained in their study ($200$~--~$1050\,{\rm d}$) does not match the range of the observed periods ($800$~--~$1400\,{\rm d}$). Based on our modelling, we infer that there are three primary sources of the difference. Firstly, the authors used [Fe/H] from $-1$ to $0$ for the whole range of masses they considered, thus introducing systems with low metallicity of $-1$ and high masses of $1.6\,M_\odot$, which are very rare in the Galaxy and result in short final periods, cf. Fig.\,\ref{fig:Galactic_FeH_Minit}. Secondly, the maximal initial mass for the progenitors was set to $1.6\,M_\odot$. As we show, higher-mass ($1.9\,M_\odot$) solar-metallicity progenitors would explain the observations of systems with periods above $1050\,{\rm d}$. Finally, the authors used several prescriptions for the angular momentum and mass loss to construct their fits, thus introducing a scatter when compared to the observed populations. With these differences accounted for, the \citet{Chen2013} model would probably result in a correlation which matches the observations more accurately.


Element diffusion in the atmospheres of sdB companions, although not included in our models, may potentially decrease their surface metallicities compared to the initial values by as much as $0.20\, {\rm dex}$ \citep{Dotter2017}. However, we expect that the material recently accreted from the red giants, chemically homogenised by convection, should have restored the initial surface metallicities of the companions. The fact that there is no visible offset between our model and the observations in Fig.\,\ref{fig:observed_period_feh} supports this idea.

A second possible observational confirmation of our model are the wide sdB binaries in the Galactic bulge. The bulge population is old and metal-rich, and the wide sdB binaries formed in this population are predicted to have on average longer orbital periods and a different slope in the $P\,-\,q$ distribution than their lower-metallicity counterparts in the Galactic thin disc (see the green versus the blue populations in Fig.\,\ref{fig:P_q_populations}). Following the reasoning in Section~\ref{sec:BinEvol}, the narrower range of final mass ratios $q$ in the $P\,-\,q$ plane may be explained by a narrower range of masses of red giants in the bulge and hence a narrower range of masses of their companions compared to the thin disc population. A different slope of the correlation compared to that of the thin disc is related to the bulge being a mono-age population with a wider metallicity spread. Observations of wide sdB binaries in the bulge are difficult as the sources are faint and located in crowded fields, and there are currently no wide sdB binaries known in the bulge.


Both of the previous tests require the orbital period of the wide sdB binaries to be known. As these systems can have periods above several years, obtaining spectroscopic observations covering the entire orbit is a very time-consuming process. One may try targeting low-metallicity candidates with shorter periods, by selecting them kinematically from the thick disc or the halo, for example through the combined use of Gaia and LAMOST data. However, our models predict that the metallicities and the masses of the companion stars should also be correlated. This relation is shown in Fig.\,\ref{fig:M2final_FeHinit}, and it does not require knowing the orbital period. The metallicity for this correlation can be obtained from a single high-resolution, high S/N spectrum. The companion mass can be determined from the surface gravity and radius or by fitting single stellar evolution models to the observed effective temperature, surface gravity, metallicity and radius \citep[see e.g.][]{Vos2018b, Maxted2015}. The radius can be obtained using the Gaia parallax, while the other parameters can be obtained from one high-resolution, high-S/N spectrum. This method only requires one observation per target system and can thus easily be used to create a large test sample for the model. We do not expect any short-period sdB binaries to pollute the sample since there are no short period sdB binaries with FGK-type companions known \citep[see e.g.][]{Kawka2015}. Similarly, we estimate chance alignments between sdBs and background or foreground MS stars to be very rare. An alternative way of obtaining the periods of sdB binaries is through high-cadence photometric variability studies, for example \citet{Otani2018}, although this method is also time-consuming.

\subsection{Gaia selection criteria}

\begin{figure}
    \includegraphics{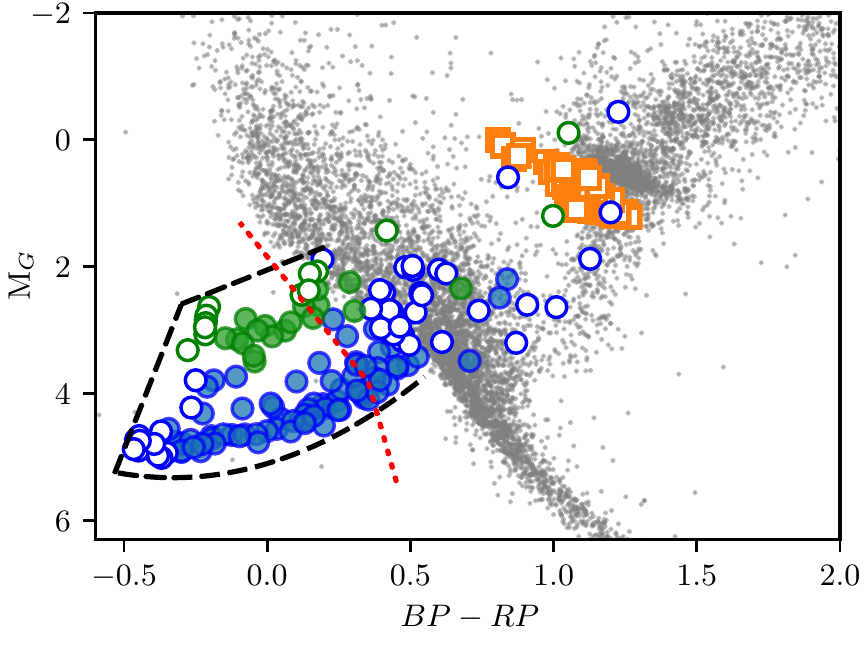}
    \caption{The Gaia colour-magnitude diagram. Full blue circles show composite-spectrum sdB+MS binaries and empty blue circles show the single-lined sdB+MS binaries. sdA+MS binaries are similarly coded in green circles. The systems that will form an sdB+MS or sdA+MS binary are shown with orange squares at the moment when the interaction phase starts. As a comparison, the Gaia colour-magnitude diagram of the Hipparcos sample is shown in grey.}
    \label{fig:Gaia_color_magnitude_diagram}
\end{figure}

Based on the effective temperature, surface gravity and radius of the MESA models, we can calculate a theoretical colour-magnitude diagram for the wide sdB binaries. The spectroscopic parameters are taken to be time-averages over the core He-burning phase. TMAP \citep{Werner2003} and Kurucz atmosphere models \citep{Kurucz1979} were used respectively for the sdB and the cool companion to calculate the synthetic magnitudes and colours. We show the resulting diagram in Fig.\,\ref{fig:Gaia_color_magnitude_diagram}. The majority of the systems are located under the main sequence and can be easily selected from a colour cut in the Gaia bands. Some wide sdB binaries, however, overlap with the main sequence. These are systems with strong MS companions that outshine the sdB and are not detectable as composite binaries (shown in open circles) as well as systems that are spectroscopically recognizable as composite sdBs. The latter systems which are not apparent from a cut in the Gaia magnitudes could still be detectable if UV photometry from, for example, GALEX \citep{Martin2003} is included. There are a few single-lined sdB binaries located at the RGB. These systems had an initial mass ratio very close to unity, and the companion has ascended the RGB during the sdB phase of the primary. The single-lined sdB+MS binaries dominated by the sdB star are located on the blue end of the diagram. These systems would photometrically and spectroscopically not look like a composite system but could be detected from RV variations.


We can define a cut in the Gaia colour-magnitude diagram that includes all composite sdB binaries. The upper and lower limits on the absolute G-band magnitude are given in the following equations:
\begin{align}
    M_G &\leq -2.06\, (G_{BP}-G_{RP})^2 - 1.35\, (G_{BP}-G_{RP}) + 5.10, \label{eq:gaia_cut_lower} \\
    M_G &\geq -1.8\, (G_{BP}-G_{RP}) + 2.05. \label{eq:gaia_cut_upper}
\end{align}
The cut in the $G_{BP}-G_{RP}$ colour is given by:
\begin{equation}
        (G_{BP}-G_{RP}) \geq -0.08\,M_G - 0.07 \label{eq:gaia_cut_color}
\end{equation}
To exclude the main sequence stars, we also apply an absolute magnitude-dependent extra colour cut, following \citet{Geier2019}, which is shown in red dotted line in Fig.\,\ref{fig:Gaia_color_magnitude_diagram}:
\begin{align}
    (G_{BP}-G_{RP}) &\leq (M_G - 1.84) / 5.6,\ \  1.0 < M_G, \leq 3.8  \label{eq:gaia_cut_geier1} \\
    (G_{BP}-G_{RP}) &\leq (M_G + 1.83) / 16.0,\ \  M_G > 3.8. \label{eq:gaia_cut_geier2}
\end{align}
After applying this main sequence Gaia cut on our simulated systems, $59$ sdB binaries and $14$ sdA binaries remain.


We show the progenitors of the sdB+MS binaries at the moment they start the mass transfer phase ($\dot{M} = 10^{-10}\,M_{\odot}/{\rm yr}$) in Fig.\,\ref{fig:Gaia_color_magnitude_diagram} in orange squares. They are all grouped just underneath the red clump but are spread broader than the typical RGB track. These predictions make it possible to search for progenitor systems just before they start the RLOF phase. Comparing the population of progenitors with the population of the long-period sdB binaries may lead to further insights into the details of mass transfer in these systems, as well as the nature of the period-eccentricity relation observed in long-period systems. The location underneath the red clump strongly reduces the number of potential candidates and makes this a more feasible observational project. The progenitors of sdB+MS binaries may potentially already be present in the current datasets, for example \citet{PriceWhelan2020}, which provide orbits of close binaries in the Galaxy.

\subsection{Galactic formation rates and local observations}

\begin{table}
\begin{center}
\begin{tabular}{ |c|c|c|c|c|c|} 
 \hline
 Type & Thin & Thick & Bulge & Halo & Total \\ 
 \hline
 $R_{\rm MW}$, ${\rm kyr}^{-1}$ & $0.91$ & $0.14$ & $0.13$ & $0.025$ & $1.2$\\
 $N_{\rm MW, tot}/10^5$ & $1.6$ & $0.31$ & $0.21$ & $0.041$ & $2.1$ \\
 $n_{\rm loc}$, ${\rm kpc}^{-3}$ & $141$ & $3.7$ & $0$ & $0.0072$ & $150$ \\
 $N_{\rm 500 pc}$ & $43$ & $1.6$ & $0$ & $0.0034$ & $45$ \\
 $N_{\rm 1 kpc}$ & $230$ & $11$ & $0$ & $0.030$ & $240$\\
 \hline
\end{tabular}
\end{center}
\caption{Formation rates and expected number counts for wide composite sdB binaries based on our MESA runs. The columns represent the Galactic components: thin disc, thick disc, the Galactic bulge, the halo and the totals. The rows show the present-day formation rate ($R_{\rm MW}$) and the expected number counts ($N_{\rm MW, tot}$) for each Galactic component, as well as the local number density ($n_{\rm loc}$) and the number of systems within $500\,{\rm pc}$ ($N_{\rm 500 pc}$) and within $1000\,{\rm pc}$ ($N_{\rm 1 kpc}$) from each Galactic component. All the numbers in the table are given to two significant digits and correspond to sdB binaries which are both identifiable observationally as wide composite sdB binaries and are included in the Gaia cuts shown in Fig.\,{\ref{fig:Gaia_color_magnitude_diagram}}. The values in this table may be rescaled to a different stellar mass of the Galaxy $M_{\rm \star, new}$ or a different average star formation rate $R_{\rm \star, new}$ over the Galactic age of $T_{\rm MW}$ by multiplying them by a factor $M_{\rm \star, new} / (6.43\cdot 10^{10}\,M_\odot)$ or $(R_{\rm \star, new}/(4.59\, M_\odot/{\textrm{yr}}))\cdot(T_{\rm MW}/(14\,\textrm{Gyr}))$, correspondingly.}
\label{Tab:GRates}
\end{table}


Our main sample of runs encloses the parameter space of all the systems formed through the Galactic history which may produce long-period composite sdB binaries degenerately at present day. Since our initial population is based on the Galactic star formation history, our set of MESA runs can be identified with the present-day formation rates of sdBs binaries, as we describe in detail in Section~\ref{sec:MainSetInit}. By applying the method to the different Galactic components, we obtain the rates shown in Tab.\,\ref{Tab:GRates}. Specifically, the present-day formation rate of all the observable wide composite sdB binaries on the main $P\,-\,q$ branch, contained in the Gaia cut (Fig.\,\ref{fig:Gaia_color_magnitude_diagram}), is $1.2\cdot10^{-3}\,{\rm yr}^{-1}$, which corresponds to $2.1\cdot 10^5$ systems currently present in the Galaxy. The primary sources of uncertainty for the numbers in Tab.\,\ref{Tab:GRates} are the Galactic stellar mass, the binary fraction and, for the $500\,{\rm pc}$ sample, the Poisson noise, which altogether may lead to an uncertainty of a factor of a few. The uncertain stellar mass of the Galaxy may be factored out from the values in Table~\ref{Tab:GRates}. In order to re-scale the values to a new Galactic stellar mass $M_{\rm \star, new}$ or a new average star formation rate $R_{\rm \star, new}$ over the Galactic age $T_{\rm MW}$, the values should be multiplied by a factor $M_{\rm \star, new} / (6.43\cdot 10^{10}\,M_\odot)$ or $(R_{\rm \star, new}/(4.59\, M_\odot/{\textrm{yr}}))\cdot(T_{\rm MW}/(14\,\textrm{Gyr}))$, correspondingly.

Despite the above uncertainty, our present-day formation rates are by $10$ to $50$ times lower than the estimate given in \citet{Han2003}. One major factor for the difference comes from the fact that \citet{Han2003} used too high star formation rate for the Galaxy. Specifically, they assumed that there is one binary with the primary more massive than $0.8\,M_\odot$ forming in the thin disc and the thick disc per year ($2$ systems per year in total) at a constant rate over $15\,\textrm{Gyr}$. In comparison, the Besanćon model we use in this study predicts that such binaries form in the whole Galaxy at a rate of $0.17\,{\textrm{yr}^{-1}}$, that is about $10$ times slower than assumed in \citet{Han2003}. Part of this difference may come from the uncertain stellar mass of the Galaxy and part of it may be since \citet{Han2003} stated that they used an upper estimate on star formation rates. 
Additionally, we selected the systems based on them being observationally identifiable as sdB binaries, which may produce another factor of $2$, as may be seen from Table~\ref{Tab:MESA_model_sets}. With these differences in mind, the rates broadly agree. We also have found the present-day formation rate for composite observable sdA binaries to be $0.27\,{\rm kyr}^{-1}$, their total Galactic number being $0.43\cdot 10^5$, subject to a similar uncertainty range of a factor of a few, as the sdB production rates. Furthermore, assuming that the mass transfer phase of the red giant progenitor systems (shown with orange squares in Fig.\,\ref{fig:Gaia_color_magnitude_diagram}) lasts on average about $10^3$--$10^4\,{\rm yr}$, typical for the MESA runs, and that the sdB phase lasts about $100\,{\rm Myr}$, we expect between several and a few tens of sdB progenitors currently in the mass transferring phase in the Galaxy.


We calculate the local number density of wide composite sdB binaries in Tab.\,\ref{Tab:GRates} by using the local stellar number densities for the solar neighbourhood for each Galactic component from \citet{Robin2003} and by assuming that the metallicity history of the stars in the solar neighbourhood is representative of Milky Way as a whole. We obtain the total local density of long-period sdB binaries of $150\,\textrm{kpc}^{-3}$. The thin disc population makes up for almost $97\,\%$ of the locally observed systems, and $3\,\%$ and $0.005\,\%$ correspond to the thick disc and the halo, respectively. The negligible fraction of the halo objects is partially due to the stellar halo containing a small fraction of the Galactic mass and partially due to the halo being much more sparse compared to the disc population. By integrating the thin, thick disc and the halo density profiles from \citet{Juric2008} and \citet{McMillan2011}, we obtain the expected number counts of the systems in the nearest $500\,{\rm pc}$ and $1000\,{\rm pc}$. We observe approximately the same fraction of thick disc objects and also a negligible fraction of halo objects in these samples. We note that while the absolute rates and numbers in Tab.~\ref{Tab:GRates} are uncertain within a factor of a few, the relative rates and number counts of sdBs in different Galactic components are much more accurate and directly reflect our knowledge about the stellar mass fractions in each of these components.

It is important to note that, while the Besanćon model was constructed to represent the Milky Way Galaxy as a whole \citep{Robin2003}, it is still, nevertheless, calibrated to provide the best representation of the observed stars in the solar neighbourhood. Our observed sample of sdB binaries and the locations of wide-period sdB binary candidates also belong to the solar neighbourhood and, therefore, should be well modelled by the age-metallicity correlation provided by the Besanćon model, also in agreement with \citet{Casagrande2011} and \citet{Bensby2014}. The Milky Way, as a whole, has a broader spread in the age-metallicity correlation compared to the solar neighbourhood, for example \citet{Feuillet2019}. While this broader spread should not have much effect on the overall Galactic rates estimated in Tab.\,\ref{Tab:GRates}, we expect the $P\,-\,q$ relation of sdB binaries to also become somewhat broader as larger parts of the Galaxy get included in the observed sample of sdB binaries.

\subsection{Constraints from/on the Galactic evolution}

We expect the metallicities of long-period sdB binaries to correlate with their vertical velocity dispersions in the Galaxy. Lower-metallicity binaries, for example, have formed earlier in the Galactic history and have experienced more dynamical interactions and thus have acquired higher vertical velocity dispersions compared to higher-metallicity longer-period sdBs which formed more recently. Based on the kinematics of white dwarfs, the vertical velocity dispersion for the oldest, lowest-metallicity sdB binaries in the thin disc is expected to be about $30\,\textrm{km/s}$ \citep{Seabroke2007} and their vertical scale height in the Galaxy is expected to be about $200\,\textrm{pc}$ \citep{Tremblay2016}. The highest-metallicity sdB binaries, instead, originate from recently-formed stars and are expected to have vertical velocity dispersions of $10\,\textrm{km/s}$ and scale heights of about $75\,\textrm{pc}$.


The radial distributions of sdB binaries in the Galaxy depend on the radial metallicity structure of the Galactic disc and, implicitly, on the radial migration history of the Galaxy and, in particular, the churning and blurring processes in the Galactic disc, for example \citet{Frankel2019} and  \citet{Feuillet2019}. The fact that the Milky Way has a radial metallicity gradient of $-0.07\,{\rm dex}/{\rm kpc}$, for example \citet{Robin2003}, suggests that lower-metallicity and shorter-period binaries should at present, on average, reside at larger galactocentric radii than higher-metallicity longer-period binaries. While the long-period sdB binary samples are too small to constrain the spatial distributions of age and metallicity in the Galaxy, it is conceivable that larger samples of exotic binaries, including short-period sdB binaries, binary white dwarfs, symbiotic binaries, cataclysmic variables and others, may in a similar manner eventually prove complementary to the current Galactic chemical evolution studies.


For example, long-period sdB binaries may be used to put constraints on the thick disc properties. The location on the $P\,-\,q$ diagram of the sdB binaries coming from the thick disc depends rather sensitively on the age assumed for the thick disc. For example, the age of $11\,\textrm{Gyr}$, assumed in \citet{Robin2003} places the thick disc binaries visibly outside of the observed $P\,-\,q$ relation, whereas the age of $10\,\textrm{Gyr}$ from the more recent study by \citet{Robin2014} lies closer to the relation, as seen in Fig.\,\ref{fig:P_q_populations}. As may be inferred from Tab.\,\ref{Tab:GRates}, constructing a complete sample of sdB stars within $1\,{\rm kpc}$ will also allow putting constraints on the Galactic mass fraction in the thick disc. For example, if the thick disc mass fraction was higher than $10$~--~$20\,\%$, the $1\,{\rm kpc}$ sample would show an overdensity of sdB binaries in the corresponding region in Fig.\,\ref{fig:P_q_populations}. We also do not expect any noticeable contribution for halo sdB stars in our observed sample, as shown in the previous section, which may be relevant to other types of sdB binaries \citep{Luo2019}.


Our study suggests that the $P\,-\,q$ relation and, generally, the binary sdB population, should be different in environments with other metallicity histories compared to the Milky Way. This concerns, in particular, dwarf galaxies, other Milky Way-like galaxies, wherein the sdB populations may potentially be probed photometrically, and especially mono-age populations such as old open and globular clusters, wherein any $P\,-\,q$ relation would have a different nature from that of the Galactic population. Similarly, our study implies that the sdB population has been rather different at different times in Galactic history. In the extreme example of the early universe, a formation channel similar to the one studied here would involve binaries formed from massive low-metallicity stars leading to bright, massive helium stars and possibly contributing to the reionisation of the universe \citep{Gotberg2017,Gotberg2020}.

\subsection{Constraining stellar evolution with sdB binaries}
\label{sec:DiscStellar}

Successful match to the observations shown in Figs.\,\ref{fig:metalicity_P_q_comparison} and \ref{fig:observed_period_feh}, lends strong support to our model of red giant mass transfer onto main sequence stars. Since all the sdB binaries in the main branch of the $P\,-\,q$ relation are the outcomes of stable mass transfer, our study serves as an additional observational confirmation that evolved red giants lead to stable mass transfer for initial mass ratios $M_{\rm RG}/M_{\rm comp}$ of up to about $1.8$, for primaries in mass range between $0.7$ and $2.0\,M_\odot$. This result is in contrast to polytropic models used, for example, in the BSE code, which predict that systems that lead to $M_{\rm sdB}/M_{\rm comp}<0.5$ should be unstable for mass ratios $M_{\rm RG}/M_{\rm comp}$ above $1$, that is always unstable in our scenario. This way, long-period sdB binaries can be used to constrain the stability of mass transfer from giant donors observationally.

Our models also validate that the red giant radii predicted by the MESA code with the default choice of the mixing length parameter are correct up to at least $\sim10\,\%$. If the actual red giant radii were smaller or larger, our model, which does not contain any fitted parameters, would lead to proportionally shorter or longer periods for all the modelled sdB binaries, which would not match the observations shown in Fig.\,\ref{fig:metalicity_P_q_comparison}.

As the observed sample becomes larger, our models can potentially provide support to the observations of anti-correlation of metallicity and the close binary fraction in the Galaxy \citep{Moe2019}. The close binary fraction at metallicities of $-0.6$ is expected to be about $50\,\%$ larger than at metallicities of $-0.2$ \citep{Moe2019}. The dependence of binary fraction on metallicity would be reflected in the relative number fractions of low-metallicity systems towards the short period end of the main branch and of the higher-metallicity systems towards the long-period end of the main branch in Fig.~\ref{fig:metalicity_P_q_comparison}.

We have used the most standard binary and Galactic evolution prescriptions with a minimal number of model parameters. Since the determining quantities for the positions of sdB binaries on the $P\,-\,q$ branch are the primary masses and the metallicities of the progenitors, we expect that the Galactic studies well calibrate these parameters. The main uncertainties of our model come from the less certain stellar mass fraction in the Galaxy, the variation of binary fractions and period distributions with metallicity and the inherent uncertainties in the standard binary mass transfer models. All these uncertainties should be reflected in the total rates of systems rather than the locations on the $P\,-\,q$ plane.

Our modelling also reinforces the idea that sdB binaries are excellent tools for constraining binary mass transfer models. While we have aimed at using the most canonical binary mass transfer model, our simulations also suggest what the effects of the assumptions about binary mass transfer may be. In particular, if we allowed the accreting star to gain significant amounts of transferred material, the resulting mass ratios would be also be significantly reduced, producing a mismatch with the observed $P\,-\,q$ relation. Since observations of the companion stars also indicate that they accrete only a few times $0.01\,M_\odot$ of material at most \citep{Vos2018b}, our study suggests that low-mass stars below about $1.5\,M_\odot$ with metallicities above $-0.5$ are generally inefficient at accreting material at mass transfer rates above $10^{-5}\,M_\odot/{\rm yr}$, which are the typical rates for sdB progenitors. 

This study does not constrain the correct angular momentum loss prescription. As discussed, for example by \citet{Rappaport1995} and \citet{Chen2013}, the formation of sdB is strongly coupled to the core mass at the end of red giant mass transfer, which in turn couples to the red giant radius at the end of mass transfer and therefore to the final orbital period. In other words, the fact the red giant core ignites to produce an sdB is coupled to the period of the final binary, rather than to the angular momentum loss history which brought the binary to that period. However, since the red giant radius at the end of mass transfer is also sensitive to [Fe/H], the final periods are directly affected by metalicity. 

Further studies of sdB binaries may help to constrain the angular momentum loss prescription. As may be seen from equation \ref{eq:PFinNonCons}, angular momentum prescription directly affects which window of initial periods may lead to the formation of sdB binaries. Furthermore, angular momentum loss prescription dictates which fraction of binaries undergo stable mass transfer. These two effects may have a direct impact on the number of systems which can lead to sdB binaries. One can also potentially probe the angular momentum loss prescription as a function of progenitor mass, by studying how the number of observed systems varies along the $P\,-\,q$ branch, although the number distribution of sdB-binaries along the $P\,-\,q$ branch is degenerate with the Galactic star formation history.  Finally, we point out that using the standard stability criteria of mass transfer was sufficient to explain the  $P\,-\,q$ relation of long-period sdB binaries. However, the choice of stability criteria may be further verified and constrained by studying through detailed MESA modelling of the populations of both short- and long-period composite sdB binaries simultaneously.

Since we have been able to explain the main branch of the $P\,-\,q$ relation through degenerate ignition, in the sample of all the binaries with the primary masses between $0.7$ and $2.0\,M_\odot$ formed through the Galactic history, the origins of the second $P\,-\,q$ branch remain unclear. It seems unlikely to us that more massive stars could produce the second branch through non-degenerate ignition. The most massive progenitors on the main branch lead to $q_{\rm final}\approx 0.3$. Along the main branch, increasing the primary mass leads to lower final mass ratios. Naively extrapolating, the more massive non-degenerately igniting stars should end up having  $q_{\rm final}\lesssim 0.3$, which is further enhanced by the fact that the final cores masses in the non-degenerate case are typically lower than in the degenerate case (unless the primary mass is larger than a few solar masses). However, a detailed binary evolution modelling of the non-degenerate channel is needed to make any conclusive predictions. It may also be possible that the second-branch objects are members of a peculiar Galactic sub-population with a metallicity of about ${\rm [Fe/H]}\approx-1$ and a broad range of ages between $5$ and $10\,{\rm Gyr}$ not included in the Besanćon model, for example \citet{Chiappini2015}, \citet{Martig2015} and \citet{Hekker2019}. However, it seems unlikely that a separate unidentified population comparable in size to that of the thin disc would be present in the solar neighbourhood. Further insights into the Galactic membership of the second branch can be gained from a kinematic analysis of the long-period sdB population. Triple stellar evolution \citep{Toonen2016} is unlikely to lead to a tight correlation seen in the second branch. Another interesting possibility could be that binary mass transfer prescriptions are also sensitive to metallicity. For example, if mass transfer were to become conservative at low metallicities, the binaries of the second branch could potentially be explained through the thick disc population, receiving their short periods due to low metallicity and the low final mass ratios due to the companion accreting additional amounts of mass. In any case, we expect that understanding the origins of the second branch will better inform us either about the Galactic populations or the binary evolution. Detecting possible systems in between the main and the second branches of the $P\,-\,q$ relation may put important constraints on some of the above scenarios.

\section{Conclusions}

In this article, we have studied the relation between the orbital periods and mass ratios observed in long-period sdB+MS binaries. This relation was discovered by \citet{Vos2018}, where it was attributed to the stability of RLOF on the red giant branch. In this work, we have shown that while the stability of RLOF will influence the range in initial mass ratios that can form an sdB star, it alone is not responsible for the strong correlation between $P$ and $q$. It is in fact the metallicity history of our Galaxy that causes the $P\,-\,q$ relation. The slow increase over time of the average metallicity in the disc, where most of the sdBs originate, causes a correlation between sdB progenitor mass and metallicity. Due to the strong dependence of the red giant radius on the metallicity, this correlation causes low-mass progenitors with lower metallicities to form sdBs at shorter orbital periods than their heavier counterparts. The fact that the sdB mass range is very narrow, combined with the narrow range of mass ratios in the original binaries which lead to sdBs, creates the strong correlation between orbital period and mass ratio. A standard binary interaction model combined with the Besanćon Galaxy evolution model can match the main branch of the observed $P\,-\,q$ relation {\rm very well}. At the same time, while there are several formation scenarios possible for the second branch of the relation, its origins remain unexplained.


Our models show a strong correlation between the final orbital period and the initial metallicity of sdB+MS binaries, which is in very good agreement with the observations. The good match with the observations serves as further support for our model. This correlation also depends on the mass of the final product and holds both for sdB+MS and He-WD+MS systems. We provide a fitting formula for these correlations in Eq.\,\ref{eq:Msdb_P_feh} and \ref{eq:Mwd_P_feh}. Furthermore, there are strong correlations between initial and observed parameters that can aid in linking observed systems to their progenitors. 


Our study also shows that it is possible to conduct a small but statistically significant population study with the MESA code. As we showed in Section \ref{sec:ResMain}, in the case of sdB binaries it is, in fact, necessary to use MESA compared to synthetic population synthesis codes such as BSE. Since presently, such codes do not capture the correct conditions for the formation of sdB stars. Using our models, we have determined the current formation rate for the long-period composite sdB binaries on the main branch of the $P\,-\,q$ relation in the Galaxy to be $1.2 \cdot 10^{-3} {\rm yr}^{-1}$, and we can expect to observe about $240$ such systems within $1\,{\rm kpc}$. Moreover, we expect the population of wide composite sdB binaries in the bulge to be different from the currently observed sample, as we show in Fig.\,\ref{fig:P_q_populations}. Our study also provides strong support for the standard mass-loss model we used for mass transfer from low-mass red giants onto main sequence stars. 


Our study has been the first to demonstrate how the chemical history of the Galaxy manifests itself in an observed binary population. We expect that using chemical evolution models of the Galaxy will have a significant impact in studying the observed binary populations, the production of which involves at least one low-mass ($M_{\rm init} < 1.6\,M_{\odot}$) component that has evolved off the main sequence. Such binary populations include hot subdwarfs, (extremely) low-mass white dwarfs, cataclysmic variables, ultra-compact X-ray binaries, symbiotic binaries, among others. In the longer term, we expect that the accurately measured properties of exotic binaries will provide new constraints on chemical evolution studies of the Galaxy.

\begin{acknowledgements}
This work was supported by a fellowship for postdoctoral researchers from the Alexander von Humboldt Foundation awarded to JV. We would like to thank Christian Sahlholdt, Diane Feuillet, Silvia Toonen, Jennifer Johnson, Chris Belczynski, Ashley Ruiter, Neige Frankel, Alina Istrate, Alejandra Romero, Ingrid Pelisoli, Paul McMillan, Astrid Lamberts, Melvyn B. Davies and Ross Church for helpful comments and discussions at different stages of this work.
\end{acknowledgements}

\bibliographystyle{aa}
\bibliography{sdBLits}

\end{document}